\documentclass[journal]{IEEEtran}

\usepackage{amsmath,amsfonts,amssymb}
\usepackage{mathrsfs}
\usepackage{algorithm}
\usepackage{algorithmic}
\usepackage{array}
\usepackage[caption=false,font=normalsize,labelfont=sf,textfont=sf]{subfig}
\usepackage{textcomp}
\usepackage{stfloats}
\usepackage{url}
\usepackage{graphicx}
\usepackage{cite}
\usepackage{multirow}
\usepackage{makecell}
\usepackage{setspace}
\usepackage{threeparttable}
\usepackage{xcolor}
\usepackage{amsthm}
\usepackage{booktabs}
\usepackage{placeins}
\usepackage{tabularx}

\newcolumntype{C}[1]{>{\centering\arraybackslash}m{#1}}
\newcolumntype{L}[1]{>{\raggedright\arraybackslash}m{#1}}

\def\BibTeX{{\rm B\kern-.05em{\sc i\kern-.025em b}\kern-.08em
    T\kern-.1667em\lower.7ex\hbox{E}\kern-.125emX}}

\begin{document}

\title{DynaSeek: Large Language Model-Driven Hybrid Dynamic Model Discovery for Grid-Tied Inverters from Limited Measurements}

\author{
Hongwei Zhen, Chao Shen, Jianfeng Niu,~\IEEEmembership{Student Member,~IEEE,}
Xin Xiang,~\IEEEmembership{Member,~IEEE,}\\
Wuhua Li,~\IEEEmembership{Senior Member,~IEEE,}
and Mingyang Sun,~\IEEEmembership{Senior Member,~IEEE}
\vspace{-30pt}
}

\markboth{Submitted to xxx}
{Zhen \MakeLowercase{\textit{et al.}}: Inverter Equation Discovery Under Partial Observation and Current Limitation}

\maketitle

\begin{spacing}{0.91}

\begin{abstract}
Discovering the unknown governing equations of grid-tied inverters from limited measurements is essential for the stability analysis of modern inverter-dominated power systems. However, existing identification methods struggle to jointly resolve unmeasured internal control memory and saturation-induced mode transitions from accessible terminal waveforms. 
To address this challenge, this paper proposes DynaSeek, a large language model (LLM)-driven framework for discovering explicit hybrid differential-algebraic equations (DAEs) of current-limited inverters. The framework integrates physics-constrained symbolic skeleton synthesis with the joint numerical estimation of shared model parameters and record-specific latent initial states across multi-island experience archives. Furthermore, two complementary structural diagnostics evaluating cross-record state closure and cross-window consistency are established to detect structural deficiencies across distinct input histories and transient stages.
Guided by diagnostic feedback, DynaSeek directs variable-dimensional latent-state expansion and current-limiting and anti-windup refinement to yield self-contained hybrid DAEs unifying unsaturated regulation, current limiting, and post-fault recovery. 
Controller-hardware-in-the-loop (CHIL) experiments and closed-loop system-level simulations demonstrate that DynaSeek reduces transient trajectory reconstruction errors by one to two orders of magnitude compared with symbolic and neural baselines while accurately preserving reference critical clearing times, enabling reliable large-signal stability assessment of inverters with inaccessible internal controls.
\end{abstract}


\section{Introduction}

\IEEEPARstart{W}{ith} the growing penetration of inverter-based resources (IBRs) in modern power systems, their multi-timescale fast control dynamics and grid interactions increasingly govern overall system stability~\cite{Nikos2020Stability,Wang2020Interaction}. Accurate characterization of inverter dynamic behavior is therefore a prerequisite for reliable transient and small-signal stability assessment. In practice, however, commercial inverters rely on proprietary control implementations whose internal architectures and parameters remain undisclosed~\cite{Sun2009Propiety}. Moreover, available measurements in practical applications are typically restricted to accessible terminal and filter quantities, leaving internal digital controller states unmeasured, including integrator states in phase-locked loops (PLLs), outer-loop regulators, and current-limiting controllers~\cite{Ramasubramanian2017Port}. This dual restriction of unknown control structures and unmeasured internal states precludes conventional white-box modeling~\cite{Wang2019DynamicBasis}. Consequently, identifying generalizable inverter dynamic models from limited measurements has become an essential requirement for stability analysis in high-IBR power grids.

Existing efforts to identify inverter dynamics from limited measurements have progressed from gray-box parameter estimation to data-driven modeling. Gray-box methods assume a predefined control structure to estimate unknown controller parameters from terminal responses, rendering their generalization sensitive to structural mismatch~\cite{Amin2019GrayBox}.
Data-driven modeling relaxes this structural prior by learning input--output dynamic mappings directly from external measurements. Early approaches used time- and frequency-domain black-box identification~\cite{Valdivia2012BlackBox} and wide-area waveform measurements~\cite{Ahmadi-Gorjayi2023PriorReduce} to capture inverter terminal responses. More recently, physics-informed neural networks have been developed to embed physical constraints into nonlinear representation learning, improving trajectory prediction accuracy and extrapolation capability under unseen disturbances~\cite{Yang2025NC,Shen2025PFNN}. The integration of latent-state learning with measurement clustering further supports multi-mode continuous-time modeling under incomplete measurements and unknown mode labels~\cite{Huang2026BlackLatent}. In addition, latent-feature-informed neural ordinary differential equations (NODEs) unify the characterization of large-signal responses and stability-relevant dynamics~\cite{Zheng2026LatentODE}. 
Nevertheless, finite training data and incomplete state measurements continue to restrict cross-condition generalization~\cite{Subedi2025Applications}. This limitation becomes particularly pronounced during deep voltage sags that trigger overcurrent limiting, where current saturation constraints abruptly reshape the active control loops~\cite{Baeckeland2024CurrentLimiting} and the continuous accumulation of internal integrator states directly governs the post-fault recovery trajectory~\cite{Bottrell2014LimitComparison}. Reliable prediction under unseen disturbances thus requires an explicit characterization of the coupling between saturation-induced mode transitions and internal state evolution.

Explicit governing-equation discovery provides an interpretable modeling approach to capture such coupled mechanisms. Sparse identification of nonlinear dynamics (SINDy) extracts parsimonious differential equations by selecting active terms from a candidate library~\cite{Brunton2016SINDy}, while physics-informed SINDy (PI-SINDy) incorporates external inputs and algebraic boundary constraints to refine power system dynamic models~\cite{Saric2021PowerSystemSINDy}. For unknown inverter controls, physics-informed sparse machine learning (PISML) couples symbolic skeletons with neural residuals and imposes local Jacobian regularization to reconcile small- and large-signal dynamics before extracting explicit control laws~\cite{Zheng2026PISML}. Whereas these approaches presume full state measurability, practical inverters with only limited measurements require integrating latent-state recovery with explicit equation identification. Autoencoder-based sparse identification addresses this requirement by simultaneously learning effective latent coordinates and their governing equations~\cite{Champion2019CoordinatesSINDy}, and joint training of hidden-state reconstruction and symbolic regression enables explicit model identification from incomplete observations~\cite{Lu2022PartialSINDy}. Even with latent-coordinate learning, the identifiable dynamic structures remain bounded by prescribed state dimensions and fixed candidate libraries. In proprietary inverters subject to saturation constraints, each additional latent state representing internal control memory requires jointly determining its driving regulation error, output coupling, and constrained evolution law during saturation. The strong coupling among these structural decisions expands the candidate search space far beyond prescribing state dimensions or pruning fixed basis functions.

Recent advances in large language models (LLMs), particularly in scientific reasoning and program synthesis, have opened new possibilities for exploring such complex structural spaces. 
Compared with conventional equation-discovery methods that rely on predefined candidate functions or expression grammars, LLMs can generate diverse candidate equations and executable programs by leveraging variable semantics, domain priors, and historical search experience. 
In scientific equation discovery, LLM-generated equation skeletons have been paired with numerical parameter estimation to iteratively refine symbolic expressions without relying on fixed function libraries~\cite{Shojaee2025llmSR}. Retrieval-augmented symbolic regression further incorporates domain knowledge and historical expressions to construct candidate building blocks for reinforcement-learning-based assembly~\cite{Guo2025SRLLM}. In power systems, LLM-based dynamic model discovery (LLM-DMD) extends this paradigm to coupled differential and algebraic equations through experience feedback and variable expansion~\cite{Shen2026LLMDMD}.
Moreover, the integration of qualitative semantic evaluation with term-wise quantitative attribution enables structural adjustments based on physical plausibility and individual term contributions~\cite{Song2026DoLQ}. However, as summarized in Table~\ref{tab:baselines_compare}, existing modeling and discovery methods still lack the combined capabilities required for current-limited inverters under limited measurements, as aggregate trajectory-fitting accuracy and general physical plausibility cannot distinguish structurally distinct candidate models that yield comparable fitting errors on recorded waveforms.

The fundamental difficulty is that limited terminal measurements reflect the combined effects of unmeasured internal control memory and saturation-induced mode transitions, making trajectory fitting alone insufficient to determine or verify the underlying model structure. 
1) \textbf{Unmeasured control memory and state non-closure:} without internal controller measurements, neither the dimension nor the input-output couplings of the latent states can be specified a priori. A candidate model lacking essential integral memory can still reduce residuals on recorded events by tuning instantaneous nonlinear terms of the accessible variables. However, when distinct past excitations produce the same candidate state but diverge under common subsequent inputs, static algebraic compensation cannot supply the missing historical information, leaving the state representation dynamically unclosed and prone to prediction drift under unseen disturbances. 
2) \textbf{Saturation-induced mode transitions and cross-stage inconsistency:} during deep voltage sags, current-limit activation and release reconfigure the active control pathways and alter internal integrator dynamics. With limiter flags and intermediate signals inaccessible, the governing saturation operators and coupled anti-windup laws remain structurally unknown. Although a model without these mechanisms can still attain low average fitting errors over clamped intervals, inaccurate constraint boundaries and integrator windup induce localized mismatches at saturation entry and post-fault release, disrupting trajectory consistency across pre-fault, current-limited, and recovery stages.

\begin{table}[t]
\centering
\scriptsize
\setlength{\tabcolsep}{1.5pt}
\renewcommand{\arraystretch}{1.2}
\caption{Comparison of Representative Inverter Modeling Methods}
\label{tab:baselines_compare}
\begin{tabular*}{\columnwidth}{
  @{\extracolsep{\fill}}
  >{\raggedright\arraybackslash}m{0.2\columnwidth}
  cccccc
  @{}
}
\toprule
\multirow{3}{*}{\textbf{Method}}
& \multicolumn{5}{c}{\textbf{Capabilities for Modeling}}
& \multirow{3}{*}{\makecell[c]{\textbf{Explicit}\\\textbf{Equations}}} \\
\cmidrule(lr){2-6}
& \makecell[c]{Limited\\Measurements}
& \makecell[c]{Library-\\Free}
& \makecell[c]{Latent\\Memory}
& \makecell[c]{Limiter \&\\Recovery}
& \makecell[c]{Structure\\Diagnostics}
& \\
\midrule
NODE~\cite{Chen2018NeuralODE}
& $\times$ & $\checkmark$ & $\times$ & $\times$ & $\times$ & $\times$ \\

Huang~\cite{Huang2026BlackLatent}
& $\checkmark$ & $\checkmark$ & $\checkmark$ & $\times$ & $\times$ & $\times$ \\

PI-SINDy~\cite{Saric2021PowerSystemSINDy}
& $\checkmark$ & $\times$ & $\times$ & $\times$ & $\times$ & $\checkmark$ \\

PISML~\cite{Zheng2026PISML}
& $\times$ & $\times$ & $\times$ & $\checkmark$ & $\checkmark$ & $\checkmark$ \\

LLM-DMD~\cite{Shen2026LLMDMD}
& $\times$ & $\checkmark$ & $\times$ & $\times$ & $\times$ & $\checkmark$ \\
\midrule
\textbf{DynaSeek}
& $\checkmark$ & $\checkmark$ & $\checkmark$ & $\checkmark$ & $\checkmark$ & $\checkmark$ \\
\bottomrule
\end{tabular*}
\end{table}

Motivated by these observations, this paper proposes DynaSeek, an LLM-driven framework that formulates hybrid differential-algebraic equation (DAE) discovery for grid-tied inverters as a diagnosis-guided progressive structural search. Starting from known filter circuit priors, DynaSeek first identifies an unsaturated baseline model by coupling LLM skeleton generation and multi-island evolutionary search with cross-record state-closure diagnosis to recover unmeasured internal control memory. In the subsequent stage, the framework transitions to large-disturbance records and leverages cross-window consistency diagnosis to augment the locked baseline structure with current-limiting operators and anti-windup recovery laws. Controller-hardware-in-the-loop (CHIL) experiments and system-level simulations confirm that the discovered hybrid DAEs accurately reproduce both unsaturated and current-limited transient trajectories while reliably predicting critical clearing times (CCTs) across diverse grid strengths and fault severities.
The main contributions of this paper are summarized as follows:

\begin{enumerate} 
    \item To the best of our knowledge, this is the first work to incorporate LLM-driven symbolic equation discovery into hybrid DAE modeling of grid-tied inverters from limited measurements. A variable-dimensional structural search framework is developed that integrates physics-constrained skeleton synthesis and multi-island experience management with the simultaneous estimation of shared model parameters and record-specific latent initial states.
    \item A state-closure-guided latent-memory identification mechanism is formulated to resolve unmeasured internal controller states. By projecting perturbed input-history sensitivities onto the null space of accessible terminal measurements, the proposed cross-record state-closure index quantifies unresolved historical effects and directs sequential variable screening and latent control-memory state augmentation.
    \item A consistency-guided two-stage expansion mechanism is developed to capture coupled current-limiting and post-fault recovery dynamics. By evaluating worst-window residuals along single-initial-state continuous trajectories without intermediate state re-initialization, the cross-window consistency index exposes localized mode-transition mismatches and guides the inherited unsaturated baseline model to incorporate current-limiting operators and anti-windup recovery laws.  
\end{enumerate}

\section{Problem Formulation}

\subsection{Hybrid Inverter Dynamics under Limited Measurements}

The transient response of a grid-tied inverter is governed by the coupled dynamics of its electrical filter stage and internal digital controller. When the proprietary controller implementation is inaccessible, terminal and filter measurements capture the external electrical response but leave the internal controller states and mode-transition logic unmeasured.

To represent these dynamics under limited measurements, the continuous state vector is partitioned into measured dynamic states $\boldsymbol{x}_{\mathrm{m}} \in \mathbb{R}^{n_{\mathrm{m}}}$ and unmeasured latent control states $\boldsymbol{\xi} \in \mathbb{R}^{n_\xi}$, where the latent dimension $n_\xi$ is unknown a priori. The measured states comprise dynamic electrical and synchronization variables directly obtained or reconstructed from available measurements, whereas the latent states represent inaccessible internal controller states, including integrator states in synchronization loops, outer-loop regulators, and current-limiting controllers. Together with the algebraic variables $\boldsymbol{z} \in \mathbb{R}^{n_z}$, the inverter dynamics under discrete operating mode $\sigma \in \Sigma$ are described by the hybrid DAE 
\begin{equation}
\begin{aligned}
\dot{\boldsymbol{x}}_{\mathrm{m}}
&=\boldsymbol{f}_{\sigma}
(\boldsymbol{x}_{\mathrm{m}},\boldsymbol{\xi},\boldsymbol{z},\boldsymbol{u},\boldsymbol{p}), \\
\dot{\boldsymbol{\xi}}
&=\boldsymbol{h}_{\sigma}
(\boldsymbol{x}_{\mathrm{m}},\boldsymbol{\xi},\boldsymbol{z},\boldsymbol{u},\boldsymbol{p}), \\
\boldsymbol{0}
&=\boldsymbol{g}_{\sigma}
(\boldsymbol{x}_{\mathrm{m}},\boldsymbol{\xi},\boldsymbol{z},\boldsymbol{u},\boldsymbol{p}),
\end{aligned}
\label{eq:hybrid_dae}
\end{equation}
where $\boldsymbol{u} \in \mathbb{R}^{n_u}$ collects known external inputs and measured terminal quantities used as model inputs, and $\boldsymbol{p}$ denotes the system parameter vector whose unknown components are estimated during model identification. The vector fields $\boldsymbol{f}_{\sigma}$ and $\boldsymbol{h}_{\sigma}$ govern the evolution of the measured dynamic states and unmeasured controller states, respectively, while $\boldsymbol{g}_{\sigma}$ defines the algebraic constraints. As the latent control state vector $\boldsymbol{\xi}$ evolves continuously, it retains essential internal control memory that shapes the subsequent trajectory of $\boldsymbol{x}_{\mathrm{m}}$. The discrete variable $\sigma$ indexes the active operating mode, with transitions among modes triggered by state-dependent transition conditions that characterize the activation and release of control constraints.

\begin{figure}[ht]
\centering
\includegraphics[width=1\linewidth]{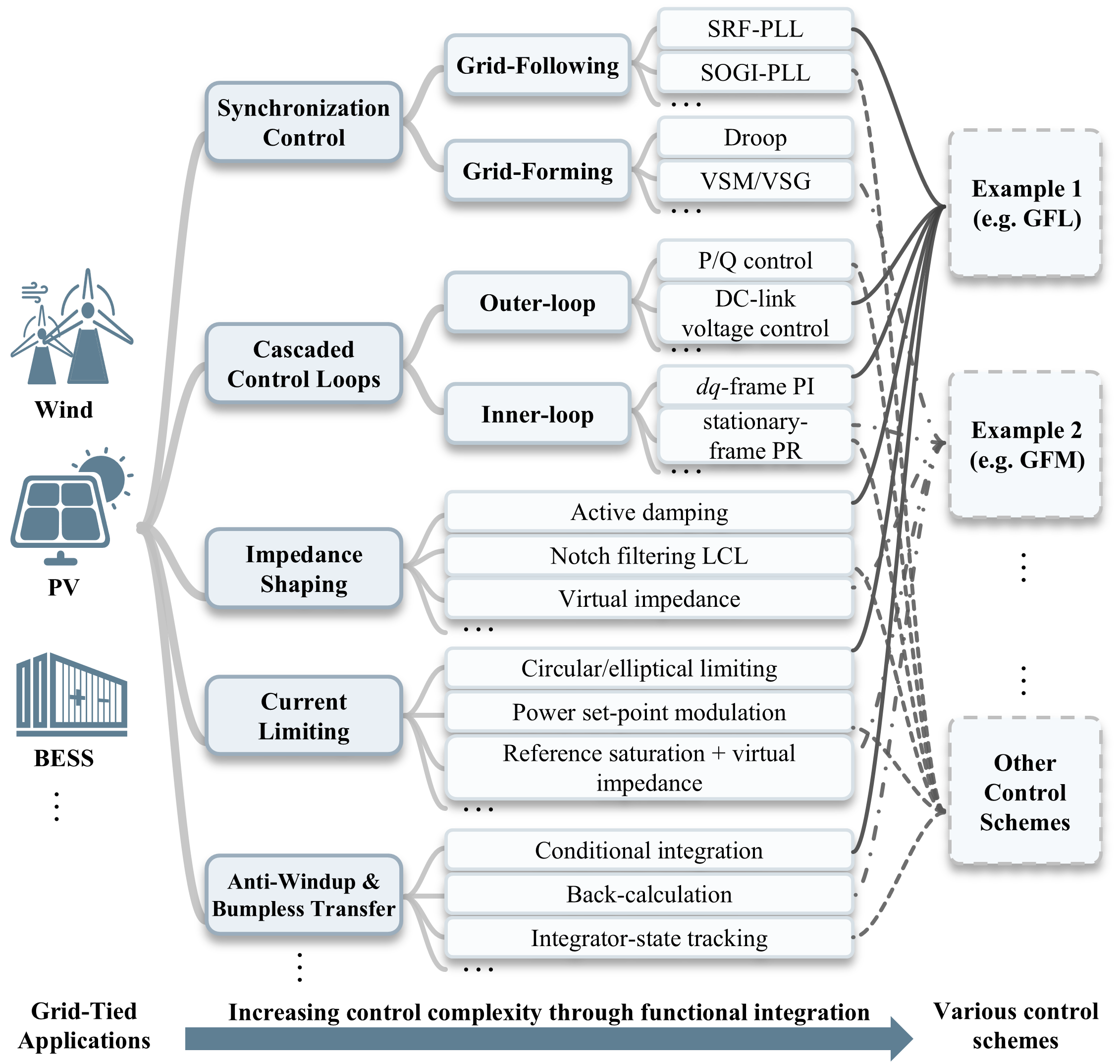}
\caption{Representative functional modules and alternative implementations in grid-tied inverters, together with illustrative combinations of these modules in practical control configurations.}
\label{fig:complex}
\end{figure}

Although the hybrid DAE provides a unified mathematical representation, practical grid-following (GFL) and grid-forming (GFM) inverter controllers admit a wide variety of structural realizations across their functional modules~\cite{Athari2017ControlReview,Rocabert2012ControlReview,Rosso2021GridForming}, as summarized in Fig.~\ref{fig:complex}. Design variations across synchronization, loop regulation, and impedance shaping directly alter both the latent dimension $n_\xi$ and the internal state dynamics $\boldsymbol{h}_{\sigma}$, whereas different current-limiting and anti-windup implementations reshape the algebraic constraints $\boldsymbol{g}_{\sigma}$ and mode-transition logic\cite{Baeckeland2024CurrentLimiting}. Owing to the cascaded and multi-loop coupling among these functional modules and the absence of internal measurements, the unknown controller dynamics cannot be reduced to a single fixed-parameter template, resulting in a high-dimensional structural hypothesis space.

\subsection{Formulation of the Equation Discovery Problem}

To formalize equation discovery over this hypothesis space, let $\mathcal{V}_{\mathrm{av}}$ denote the fixed information set compiled from synchronized input and measurement trajectories across multiple disturbance events, and let $\mathcal{D}_{\mathrm{tr}}$ denote the collection of training records. Each record in $\mathcal{D}_{\mathrm{tr}}$ spans a complete disturbance transient whose measured outputs comprise the measured states $\boldsymbol{x}_{\mathrm{m}}$ and accessible algebraic-interface quantities. Given the known electrical laws, the discovery task is to determine a latent state representation, symbolic governing relations constructed from $\mathcal{V}_{\mathrm{av}}$ and $\boldsymbol{\xi}$, and associated numerical parameters that reproduce the observed transient responses across $\mathcal{D}_{\mathrm{tr}}$.

Specifically, each candidate structure $S_j$ defines a latent state dimension $n_\xi$ together with the symbolic forms of the differential vector fields $\boldsymbol{f}_{\sigma}$ and $\boldsymbol{h}_{\sigma}$ and the algebraic constraint mapping $\boldsymbol{g}_{\sigma}$. Its parameter vector $\boldsymbol{\theta}_j$ collects the unknown coefficients in $\boldsymbol{p}$ shared across training records, whereas $\boldsymbol{\xi}_{0,j,i}$ represents the initial condition of the unmeasured control states for record $i$. Stacking these initial-state vectors as columns of $\boldsymbol{\Xi}_{0,j}$ yields the candidate-specific parameter and initial-state estimation problem
\begin{equation}
\left(
\widehat{\boldsymbol{\theta}}_j,
\widehat{\boldsymbol{\Xi}}_{0,j}
\right)
\in
\underset{\substack{
\boldsymbol{\theta}_j\in\Theta_j\\
\boldsymbol{\Xi}_{0,j}\in\mathcal X_{0,j}}}{\operatorname{arg\,min}}
L_{{\rm fit},j}
\left(
\boldsymbol{\theta}_j,
\boldsymbol{\Xi}_{0,j}
\mid
\mathcal D_{\rm tr}
\right),
\label{eq:parameter_estimation}
\end{equation}
where $\Theta_j$ and $\mathcal{X}_{0,j}$ denote the feasible parameter and initial-condition sets governed by physical bounds and steady-state consistency requirements, respectively, and $L_{\mathrm{fit},j}$ quantifies the trajectory-fitting error on $\mathcal{D}_{\mathrm{tr}}$. While this parameter and initial-state estimation problem determines the shared coefficients and record-specific initial conditions for a fixed structure $S_j$, identifying the symbolic structure $S_j$ itself across the variable-dimensional candidate space constitutes the core objective of the discovery problem.

Beyond fitting the training records, an admissible candidate structure $S_j$ needs to satisfy dimensional homogeneity and maintain solvable algebraic equations across both unsaturated and current-limited operating modes, ensuring well-posed time-domain simulation. 
Furthermore, when predicting an unseen disturbance, the model determines its initial latent state solely from a finite pre-disturbance window and then integrates forward autonomously under the prescribed inputs.
Overall, the discovery task aims to identify a concise hybrid DAE that accurately predicts unseen transient responses while maintaining dynamic consistency across operating modes.

\section{Methodology}

\begin{figure*}[t]
\centering
\includegraphics[width=1\linewidth]{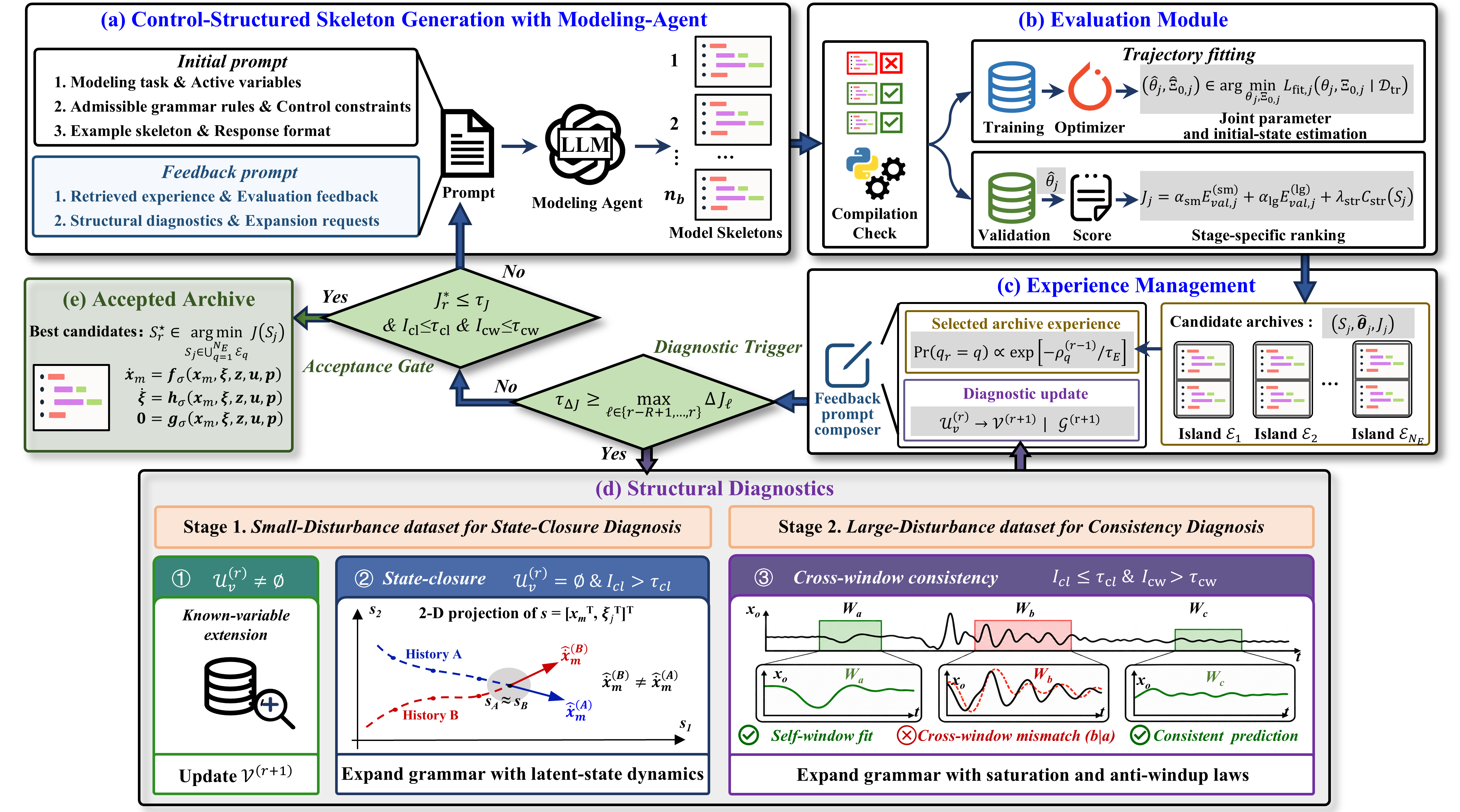}
\vspace{-1em}
\caption{Overview of the proposed DynaSeek framework, integrating LLM-based skeleton generation, joint parameter and initial-state estimation, multi-island experience management, and structural diagnostics. During diagnosis, Stage 1 transitions to Stage 2 once $J_r^\star\leq\tau_J^{(\mathrm{sm})}$ and $I_{\mathrm{cl}}\leq\tau_{\mathrm{cl}}$ hold for $R$ consecutive rounds. The accepted unsaturated structure is then carried into Stage 2, with the objective switching from $J^{(\mathrm{sm})}$ to $J^{(\mathrm{lg})}$ and the acceptance threshold from $\tau_J^{(\mathrm{sm})}$ to $\tau_J^{(\mathrm{lg})}$.}
\label{fig:Framework}
\end{figure*}

To overcome the structural complexity of governing-equation discovery from limited measurements and the inadequacy of aggregate trajectory fitting in exposing structural deficiencies, this section presents the DynaSeek framework.
As illustrated in Fig.~\ref{fig:Framework}, DynaSeek integrates LLM-driven symbolic skeleton synthesis, joint parameter and initial-state estimation, and multi-island experience management within a closed-loop diagnostic architecture. Guided by two complementary structural diagnostics evaluating cross-record state closure and cross-window consistency, the framework progressively refines the latent control-memory states and mode-dependent governing laws to discover an explicit hybrid DAE capturing unsaturated operation, current limiting, and post-fault recovery.

\subsection{Model Generation and Parameter Estimation}
\label{subsec:model_generation}
\begin{figure}[ht]
\centering
\includegraphics[width=0.95\linewidth]{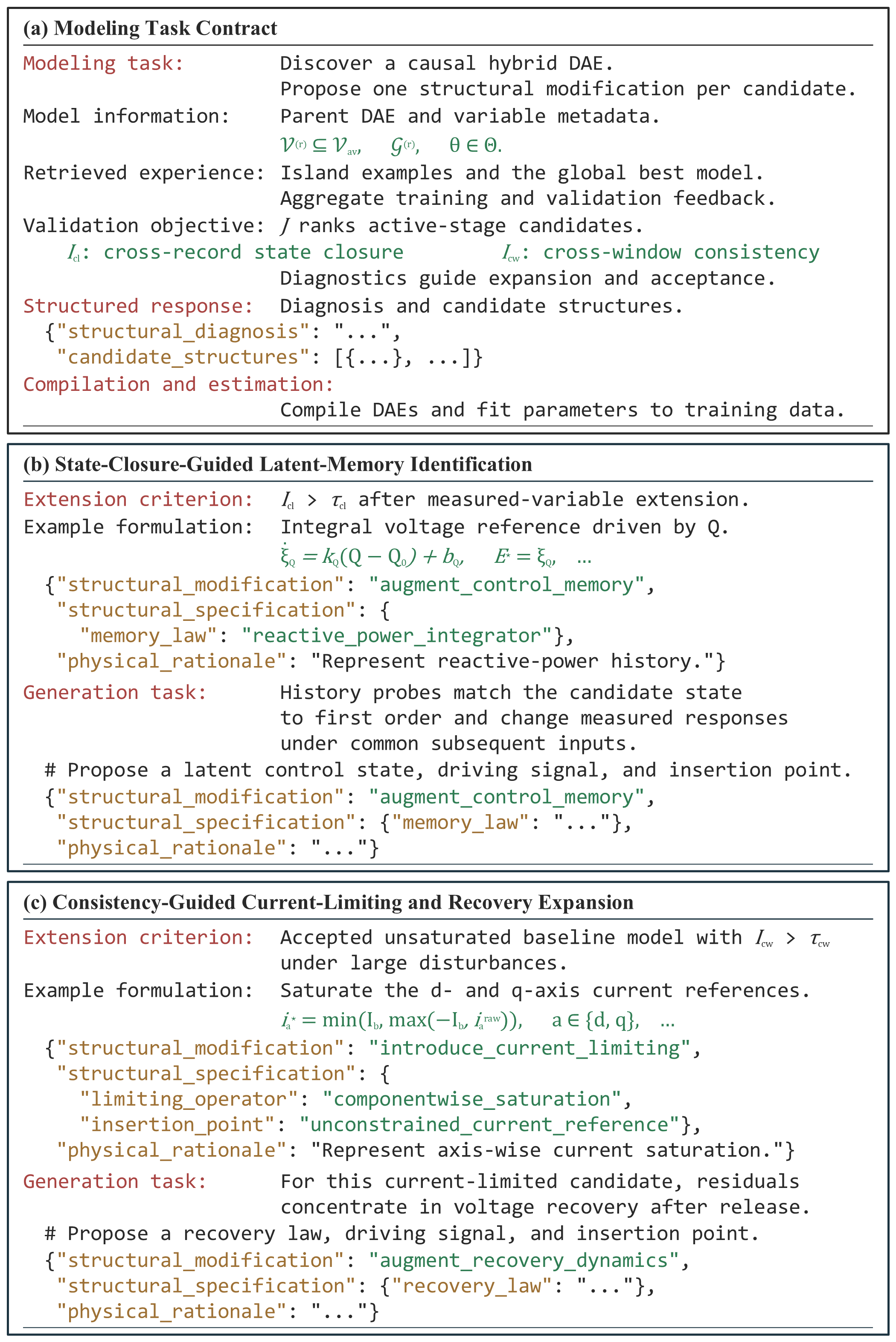}
\caption{Prompt design for the Modeling Agent.}
\label{fig:prompt_design}
\end{figure}

At each search round $r$, an LLM-based modeling agent translates accessible terminal and filter quantities together with physical priors into candidate equation skeletons, as depicted in Fig.~\ref{fig:prompt_design}. The modeling task is specified through a structured prompt $\Pi^{(r)}$ comprising an active variable subset $\mathcal{V}^{(r)}$, an admissible structural grammar $\mathcal{G}^{(r)}$, retrieved archive examples, and diagnostic feedback, while variable metadata encode physical roles, dimensions, units, and signal sources. Conditioned on the structured prompt at sampling temperature $T$, the modeling agent generates a batch of $n_{\mathrm{b}}$ candidate skeletons $S_j^{(r)}$:
\begin{equation}
    S_{j}^{(r)}
    \sim
    p_{\mathrm{LLM}}
    \left(
        \cdot \mid \Pi^{(r)}, T
    \right),
    \qquad j=1,\ldots,n_{\mathrm{b}}.
    \label{eq:candidate_generation}
\end{equation}
Each generated skeleton specifies the symbolic forms of the differential vector fields and algebraic constraints, the parameter bounds, the required variables, and candidate saturation locations along the cascaded control loops permitted by the active grammar. During local compilation, each skeleton is verified against the active variable set, the admissible grammar, dimensional homogeneity, and algebraic solvability to instantiate an executable hybrid DAE.

To estimate the shared parameters and initial latent states on the training dataset $\mathcal{D}_{\mathrm{tr}}$, each compiled hybrid DAE is integrated forward under the recorded training inputs from shared model parameters $\boldsymbol{\theta}_j$ and record-specific initial latent states $\boldsymbol{\xi}_{0,j,i}$, with the algebraic variables resolved internally. At each time sample $k$ of training record $i$, the discrepancy between the measured output vector $\boldsymbol{y}_{i,k} \in \mathbb{R}^{n_y}$ and the model-predicted output vector $\hat{\boldsymbol{y}}_{j,i,k}$ defines the normalized prediction residual:
\begin{equation}
    \boldsymbol{e}_{j,i,k}
    =
    \boldsymbol{D}_y^{-1}
    \left(
        \boldsymbol{y}_{i,k}
        -
        \hat{\boldsymbol{y}}_{j,i,k}
    \right).
    \label{eq:normalized_error}
\end{equation}
Here, the measured output vector collects the measured dynamic states and accessible algebraic-interface quantities, while the positive diagonal scaling matrix $\boldsymbol{D}_y$ normalizes the output channels across the training dataset. To quantify waveform alignment across all $N_{\mathrm{tr}}$ training records of sample length $K_i$, the trajectory-fitting error $L_{\mathrm{fit},j}$ is formulated as:
\begin{equation}
    L_{\mathrm{fit},j}
    =
    \frac{1}{N_{\mathrm{tr}}}
    \sum_{i=1}^{N_{\mathrm{tr}}}
    \frac{1}{K_i}
    \sum_{k=1}^{K_i}
    \left\|
        \boldsymbol{e}_{j,i,k}
    \right\|_{\boldsymbol{W}_y}^{2}.
    \label{eq:fitting_objective}
\end{equation}
The fixed positive semidefinite matrix $\boldsymbol{W}_y$ weights the relative contribution across output channels, and minimizing this trajectory-fitting objective jointly yields the estimated parameter vector $\hat{\boldsymbol{\theta}}_j$ and the training initial-state matrix $\hat{\boldsymbol{\Xi}}_{0,j}$.

Unlike training records where initial states are jointly estimated across the full transient, evaluating a fitted candidate on an unseen record with fixed parameters $\hat{\boldsymbol{\theta}}_j$ requires inferring the initial latent state $\hat{\boldsymbol{\xi}}_{0,j,i}$ solely from the recorded input-output samples over a pre-disturbance window of length $K_{\mathrm{pre}}$ through the initialization operator $\mathcal{I}_j$:
\begin{equation}
    \widehat{\boldsymbol{\xi}}_{0,j,i}
    =
    \mathcal{I}_j
    \left(
        \left\{
            \boldsymbol{u}_{i,k},
            \boldsymbol{y}_{i,k}
        \right\}_{k=1}^{K_{\mathrm{pre}}};
        \widehat{\boldsymbol{\theta}}_j
    \right).
    \label{eq:initialization_map}
\end{equation}
Starting from an algebraic steady-state estimate derived from accessible terminal and filter quantities, this operator solves a bounded regularized least-squares problem over the finite pre-disturbance window and propagates the latent state to the disturbance onset, from which the hybrid DAE integrates forward autonomously over the post-disturbance prediction horizon.

\subsection{Model Evaluation and Experience Management}
\label{subsec:model_evaluation}

Building upon pre-disturbance initial-state inference and subsequent autonomous integration, fitted candidates are evaluated on validation records strictly disjoint from the training set to select effective control structures for retention and evolutionary reuse. To separate continuous loop regulation from current-limiting dynamics, the validation records are partitioned into a small-disturbance dataset $\mathcal{D}_{\mathrm{sm,val}}$ for unsaturated operation and a large-disturbance dataset $\mathcal{D}_{\mathrm{lg,val}}$ for current-limited operation involving constraint activation and post-fault recovery. Evaluating all candidates over identical pre-disturbance windows, terminal electrical and synchronization output channels, and normalization scales keeps structures with distinct latent dimensions directly comparable. 

Combining the mean weighted quadratic validation errors $E_{\mathrm{val},j}^{(\mathrm{sm})}$ and $E_{\mathrm{val},j}^{(\mathrm{lg})}$ over the two datasets with a structural complexity functional $C_{\mathrm{str}}(S_j)$, the composite candidate score is defined as:
\begin{equation}
J(S_j)
=
\alpha_{\mathrm{sm}} E_{\mathrm{val},j}^{(\mathrm{sm})}
+
\alpha_{\mathrm{lg}} E_{\mathrm{val},j}^{(\mathrm{lg})}
+
\lambda_{\mathrm{str}} C_{\mathrm{str}}(S_j),
\label{eq:validation_score}
\end{equation}
where $\alpha_{\mathrm{sm}}$ and $\alpha_{\mathrm{lg}}$ denote stage-specific validation weights, $\lambda_{\mathrm{str}}$ is the complexity penalty weight, and $C_{\mathrm{str}}(S_j)$ denotes a normalized structural count of selected variable groups, latent memory states, equation terms, and control features. By switching $\alpha_{\mathrm{sm}}$ and $\alpha_{\mathrm{lg}}$ between the two discovery stages, candidate ranking yields $J^{(\mathrm{sm})}$ during unsaturated-mode discovery to isolate the baseline regulation loops, and transitions to $J^{(\mathrm{lg})}$ during large-disturbance discovery to jointly evaluate current-limiting and recovery laws while preserving unsaturated accuracy.

As practical inverter controllers admit diverse realizations across synchronization, outer-loop regulation, and current-limiting modules, relying on a single model pool risks premature convergence to suboptimal equation structures. To preserve structural diversity across alternative control hypotheses, DynaSeek maintains $N_{\mathrm{E}}$ persistent experience islands $\mathscr{E}_1,\ldots,\mathscr{E}_{N_{\mathrm{E}}}$, each initialized with a common baseline structural seed. Prior to round $r$, the islands are ranked from best to worst by their lowest active-stage score achieved over the preceding search history to assign normalized ranks $\rho_q^{(r-1)}$. The active experience island $q_r$ for round $r$ is then sampled via the Boltzmann distribution:
\begin{equation}
\Pr(q_r=q)
=
\frac{\exp[-\rho_q^{(r-1)}/\tau_{\mathrm{E}}]}
{\displaystyle\sum_{\ell=1}^{N_{\mathrm{E}}}
\exp[-\rho_\ell^{(r-1)}/\tau_{\mathrm{E}}]},
\label{eq:island_sampling}
\end{equation}
where the positive temperature parameter $\tau_{\mathrm{E}}$ balances exploitation of high-performing islands with continued exploration across alternative structural branches.

Within the sampled island, the best-performing archived model is retrieved as the in-context island example and paired with the global-best model and its diagnostic summary in the structured prompt, enabling cross-island structural recombination. Following batch evaluation, all newly evaluated candidates alongside their estimated parameters, validation scores, diagnostic feedback, and parent lineages are archived back into the active island to preserve within-island evolutionary lineages, and the global-best structure $S_r^*$ and corresponding score $J_r^*$ across all islands are updated as:
\begin{equation}
S_r^\star
\in
\underset{S_j\in\bigcup_{q=1}^{N_{\mathrm{E}}}\mathscr{E}_q}
{\operatorname{arg\,min}}\, J(S_j),
\qquad
J_r^\star=J(S_r^\star).
\label{eq:global_best}
\end{equation}

While evolutionary search within a fixed hypothesis space refines symbolic expressions, missing measured variables, latent control states, or current-limiting and anti-windup laws eventually cause the global-best score to stagnate or reach the error target while retaining unresolved structural mismatches. Accordingly, stage acceptance requires jointly satisfying the validation score target $J_r^* \le \tau_J$ and the stage-specific structural diagnostic criterion, where $\tau_J$ takes the value $\tau_J^{(\mathrm{sm})}$ in the unsaturated stage and $\tau_J^{(\mathrm{lg})}$ in the large-disturbance stage. Whenever either acceptance condition remains unsatisfied, DynaSeek tracks the single-round score improvement $\Delta J_l = J_{l-1}^* - J_l^*$ and initiates structural diagnosis upon detecting score stagnation over $R$ consecutive rounds:
\begin{equation}
\tau_{\Delta J}
\geq 
\max_{\ell\in\{r-R+1,\ldots,r\}}\Delta J_{\ell},
\label{eq:diagnostic_trigger}
\end{equation}
where the positive threshold $\tau_{\Delta J}$ specifies the minimum meaningful score improvement. Satisfying this stagnation criterion activates the two complementary structural diagnostics detailed in the following subsections to sequentially guide latent-memory identification and current-limiting expansion.

\subsection{State-Closure-Guided Latent-Memory Identification}
\label{subsec:state_closure}

During unsaturated operation, a candidate model that omits essential integrator states may still achieve low trajectory errors on recorded events by tuning static nonlinear terms of the accessible terminal variables. To expose historical input effects left uncaptured by the candidate state vector, DynaSeek performs cross-record state-closure diagnosis during unsaturated-mode discovery.

By local observability theory~\cite{Massiani2023Observability}, a dynamically closed state representation guarantees that input-history variations leaving the current state invariant to first order also leave the subsequent forward response invariant. DynaSeek evaluates this invariance property by superimposing smooth probe waveforms onto the external input vector over a finite pre-evaluation interval:
\begin{equation}
\boldsymbol{u}_{\mathrm{e}}(t;\boldsymbol{a})
=
\boldsymbol{u}_{\mathrm{e},\mathrm{b}}(t)
+
\sum_{\ell=1}^{N_{\mathrm{p}}}
(a_\ell-a_{\mathrm{b},\ell})\boldsymbol{\phi}_{\ell}(t),
\label{eq:hr_history_inputs}
\end{equation}
where $\boldsymbol{u}_{\mathrm{e},\mathrm{b}}(t)$ denotes the baseline input trajectory, $N_{\mathrm{p}}$ is the number of superimposed probes, $\boldsymbol{\phi}_\ell(t)$ represents the $\ell$-th smooth probe waveform, and $\boldsymbol{a} = [a_1, \dots, a_{N_{\mathrm{p}}}]^{\mathsf{T}}$ collects the probe amplitudes perturbed around the baseline coefficients $\boldsymbol{a}_{\mathrm{b}}$. All probing trials initialize from an identical steady-state operating point, and the probe waveforms decay smoothly to zero at the evaluation instant $t_{\mathrm{e}}$ so that all trials share an identical external input trajectory thereafter. 

At the evaluation instant, let $\boldsymbol{w}(t_{\mathrm{e}}; \boldsymbol{a})$ collect the measured dynamic states and accessible algebraic-interface quantities recorded from the inverter. Propagating the candidate latent dynamics of structure $S_j$ over each perturbed input history under the estimated parameters yields the composite candidate state vector $\boldsymbol{s}_j(\boldsymbol{a})$:
\begin{equation}
\boldsymbol{s}_j(\boldsymbol{a})
=
\begin{bmatrix}
\boldsymbol{w}(t_{\mathrm{e}};\boldsymbol{a})\\
\boldsymbol{\xi}_j(t_{\mathrm{e}};\boldsymbol{a})
\end{bmatrix}.
\label{eq:hr_current_representation}
\end{equation}

To quantify the subsequent system evolution, the reference forward response $\boldsymbol{y}_{\mathrm{f}}(\boldsymbol{a})$ is constructed from the finite-horizon state increments $\Delta_\tau \boldsymbol{x}_{\mathrm{m}}(\boldsymbol{a}) = \boldsymbol{x}_{\mathrm{m}}(t_{\mathrm{e}} + \tau; \boldsymbol{a}) - \boldsymbol{x}_{\mathrm{m}}(t_{\mathrm{e}}; \boldsymbol{a})$ across a set of prescribed horizons $\tau \in \mathcal{T}$:
\begin{equation}
\boldsymbol{y}_{\mathrm{f}}(\boldsymbol{a})
=
\left[
\boldsymbol{D}_{\mathrm{f}}^{-1}
\Delta_{\tau}\boldsymbol{x}_{\mathrm{m}}(\boldsymbol{a})/\tau
\right]_{\tau\in\mathcal{T}},
\label{eq:hr_forward_response}
\end{equation}
where the positive diagonal scaling matrix $\boldsymbol{D}_{\mathrm{f}}$ normalizes the finite-horizon state increments across channels.   

Applying centered finite-difference perturbations around $\boldsymbol{a}_{\mathrm{b}}$ yields the local sensitivity matrices $\boldsymbol{G}_w$ and $\boldsymbol{G}_{\mathrm{f}}$ of the accessible quantities and reference forward response, together with the candidate latent sensitivity matrix $\boldsymbol{G}_{\xi,j}$. Expressed in energy-normalized probe coordinates, a nonzero history perturbation direction $\boldsymbol{v}$ leaves the composite candidate state invariant to first order under the null-space relation:
\begin{equation}
\boldsymbol{G}_{s,j}\boldsymbol{v}
=
\boldsymbol{0},
\qquad
\boldsymbol{G}_{s,j}
=
\begin{bmatrix}
\boldsymbol{G}_w\\
\boldsymbol{G}_{\xi,j}
\end{bmatrix}.
\label{eq:hr_joint_matching}
\end{equation}

By the chain rule, unique determination of the subsequent forward response from the composite candidate state under identical future inputs requires the local sensitivity null spaces to satisfy the first-order state-closure condition:
\begin{equation}
\ker(\boldsymbol{G}_{s,j})
\subseteq
\ker(\boldsymbol{G}_{\mathrm{f}}),
\label{eq:hr_closure_condition}
\end{equation}
where $\ker(\cdot)$ denotes the matrix null space. Any violation of this null-space inclusion exposes a dynamically relevant input-history direction that alters the future response despite remaining indistinguishable under the candidate state representation.

To isolate the contribution of the latent control-memory vector, let the matrix $\boldsymbol{N}_0$ form an orthonormal basis for the null space of $\boldsymbol{G}_w$, and project the forward-response and latent sensitivity matrices onto this subspace:
\begin{equation}
\begin{aligned}
\boldsymbol{G}_w\boldsymbol{N}_0
&=\boldsymbol{0},\\
\boldsymbol{R}
&=\boldsymbol{G}_{\mathrm{f}}\boldsymbol{N}_0,\\
\boldsymbol{H}_j
&=\boldsymbol{G}_{\xi,j}\boldsymbol{N}_0.
\end{aligned}
\label{eq:hr_conditional_responses}
\end{equation}
Geometrically, $\boldsymbol{N}_0$ spans input-history perturbations that produce identical accessible terminal quantities at the evaluation instant. Satisfying state closure within this subspace requires the row space of $\boldsymbol{R}$ to lie within the row space of $\boldsymbol{H}_j$, which defines the cross-record state-closure index:
\begin{equation}
I_{\mathrm{cl}}(S_j)
=
\left\|
\boldsymbol{V}_R\boldsymbol{N}_{H,j}
\right\|_2,
\label{eq:hr_closure_index}
\end{equation}
where the matrix $\boldsymbol{N}_{H,j}$ forms an orthonormal basis for the null space of $\boldsymbol{H}_j$, and $\boldsymbol{V}_R$ contains an orthonormal row basis for the effective row space of $\boldsymbol{R}$ extracted through singular value decomposition under a unified rank-resolution threshold~\cite{Lerman2018Subspace}. 
Bounded between zero and unity, the index $I_{\mathrm{cl}}$ quantifies the largest unresolved historical memory component within the probed perturbation subspace, where values exceeding $\tau_{\mathrm{cl}}$ expose first-order state-closure violations.

An elevated cross-record state-closure index above the threshold $\tau_{\mathrm{cl}}$ directs DynaSeek to first screen the available measurement set $\mathcal{V}_{\mathrm{av}}$ according to dimensional homogeneity and input-output causality, incorporating the eligible unused variable subset $\mathcal{U}_{\mathrm{v}}^{(r)}$ into the active variable set:
\begin{equation}
\mathcal{V}^{(r+1)}
=
\mathcal{V}^{(r)}\cup\mathcal{U}_{\mathrm{v}}^{(r)}.
\label{eq:variable_set_update}
\end{equation}

Once all eligible measured variables have been incorporated, persistent state-closure violations drive the augmentation of unmeasured internal controller states through latent control states $\xi_{j,h}$ governed by first-order integral or damped-integral dynamics:
\begin{equation}
\dot{\xi}_{j,h}
=
k_{j,h}\epsilon_{j,h}^{\mathrm{ctrl}}
-
a_{j,h}\xi_{j,h},
\label{eq:control_memory_dynamics}
\end{equation}
where $k_{j,h}$ denotes the integral gain of the $h$-th added coordinate, the nonnegative coefficient $a_{j,h}$ determines the damping rate with zero yielding pure integral action, and $\epsilon_{j,h}^{\mathrm{ctrl}}$ represents a dimensionally consistent tracking-error signal constructed within the corresponding control loop. The newly introduced coefficients are jointly estimated with the existing model parameters on the training dataset. 

Achieving state closure prior to reaching the small-disturbance score target keeps the structural search within the current state dimension to refine the symbolic governing equations. The unsaturated baseline model is established by simultaneously satisfying the unsaturated accuracy and state-closure criteria:
\begin{equation}
J^{(\mathrm{sm})}(S_j)\leq\tau_J^{(\mathrm{sm})},
\qquad
I_{\mathrm{cl}}(S_j)\leq\tau_{\mathrm{cl}}.
\label{eq:smooth_acceptance}
\end{equation}
Maintaining both criteria over $R$ consecutive confirmation rounds locks the global-best candidate as the unsaturated baseline model $S_{\mathrm{sm}}^*$, whose governing equations, estimated parameters, and initialization operator are inherited by the subsequent large-disturbance stage.

\subsection{Consistency-Guided Current-Limiting and Recovery Expansion}
\label{subsec:consistency_expansion}

With the unsaturated baseline model locked, DynaSeek resets the score history, retains the source-island identity, and transitions to large-disturbance discovery under the combined validation score. During current-limited voltage sags, low average residuals over quasi-steady-state or clamped intervals can mask localized mismatches during saturation entry and post-fault anti-windup recovery. To verify consistent trajectory evolution across successive transient stages independently of pre-disturbance initialization errors, DynaSeek performs cross-window consistency diagnosis~\cite{Schoukens2019Nonlinear} during large-disturbance discovery.

In this diagnosis, each large-disturbance record in the diagnostic validation set $\mathcal{D}_{\mathrm{lg,val}}$ is integrated forward as a single continuous trajectory under the fixed parameter vector from a single record-specific initial latent state $\boldsymbol{\xi}_{0,j,i}$. All local evaluation windows are extracted directly from this continuous trajectory without intermediate state re-initialization. Under the training normalization scales, the window-level root-mean-square residual across all $n_y$ measured output channels is expressed as:
\begin{equation}
\ell_{j,i,w}(\boldsymbol{\xi}_{0,j,i}) 
= \left( \frac{1}{N_{i,w} n_y} \sum_{k \in \mathcal{K}_{i,w}^{\mathrm{diag}}} \Vert{} \boldsymbol{e}_{j,i,k}(\boldsymbol{\xi}_{0,j,i}) \Vert{}_2^2 \right)^{1/2},
\label{eq:diagnostic_window_error}
\end{equation}
where $w \in \mathcal{W}_i$ indexes the local evaluation windows of diagnostic record $i$, and $\mathcal{K}_{i,w}^{\mathrm{diag}}$ collects the $N_{i,w}$ time samples within window $w$.   

Record-level diagnostic evaluation solves a multistart minimax optimization problem over the feasible initialization set $\mathcal{X}_{0,j,i}^{\mathrm{diag}}$ to identify the single initial latent state that minimizes the worst-window residual along the continuous trajectory:
\begin{equation}
E_{j,i}^{\mathrm{cw}} 
= \min_{\boldsymbol{\xi}_{0,j,i} \in \mathcal{X}_{0,j,i}^{\mathrm{diag}}} \max_{w \in \mathcal{W}_i} \ell_{j,i,w}(\boldsymbol{\xi}_{0,j,i}).
\label{eq:record_consistency_error}
\end{equation}
In this minimax formulation, non-convergent or numerically infeasible trajectories are assigned an infinite error. Taking the maximum record-level error across the diagnostic validation set defines the cross-window consistency index:
\begin{equation}
I_{\mathrm{cw}}(S_j)
=
\max_{i\in\mathcal{D}_{\mathrm{lg,val}}}
E_{j,i}^{\mathrm{cw}}.
\label{eq:cross_window_index}
\end{equation}
Consequently, full-record initial-state optimization isolates structural inconsistencies across pre-fault, current-limited, and post-fault windows during diagnosis, whereas predictive validation relies solely on the finite pre-disturbance observation window.

A cross-window consistency index exceeding the threshold $\tau_{\mathrm{cw}}$ triggers the expansion of the admissible structural grammar to incorporate current-limiting operators, including magnitude saturation and radial projection operators, together with their coupled anti-windup recovery laws. Specifically, a parameterized saturation operator $\mathcal{O}_m$ maps an unconstrained reference $\zeta_{\mathrm{raw}}$ to its saturated reference $\zeta_{\mathrm{con}}$:
\begin{equation}
\zeta_{\mathrm{con}} = \mathcal{O}_m(\zeta_{\mathrm{raw}}, \boldsymbol{\theta}_{\mathrm{op},m}),
\label{eq:constraint_operator}
\end{equation}
Here, the vector $\boldsymbol{\theta}_{\mathrm{op},m}$ collects the associated threshold or projection parameters appended to the candidate parameter vector $\boldsymbol{\theta}_j$, while the corresponding anti-windup recovery terms are incorporated into the latent state and algebraic equations to characterize constraint propagation across the cascaded control loops. Both the inherited and newly introduced parameters are jointly estimated on the training set and held fixed during subsequent validation and diagnostic evaluation.   

Final model acceptance in the large-disturbance stage retains the unsaturated acceptance criteria and additionally enforces the large-disturbance accuracy and consistency criteria:
\begin{equation}
J^{(\mathrm{lg})}(S_j)\leq\tau_J^{(\mathrm{lg})},
\qquad
I_{\mathrm{cw}}(S_j)\leq\tau_{\mathrm{cw}}.
\label{eq:hybrid_acceptance}
\end{equation}
Satisfying both the unsaturated and large-disturbance acceptance criteria over $R$ consecutive confirmation rounds under fixed model parameters establishes a self-contained hybrid DAE capturing unsaturated operation, current limiting, and post-fault recovery. The complete discovery procedure is summarized in Algorithm~\ref{alg:diagnostic_routing}.

\begin{algorithm}[t!]
\caption{Diagnosis-Guided Hybrid Dynamic Model Discovery}
\label{alg:diagnostic_routing}
\footnotesize
\begin{algorithmic}[1]
\REQUIRE Training set $\mathcal{D}_{\mathrm{tr}}$, validation sets $\mathcal{D}_{\mathrm{sm,val}}$ and $\mathcal{D}_{\mathrm{lg,val}}$, initial variable set $\mathcal{V}^{(0)}$, initial grammar $\mathcal{G}^{(0)}$, and baseline structural seed.
\ENSURE Discovered hybrid DAE $(S^*, \hat{\theta}^*, \mathcal{I}^*)$.
\STATE Initialize islands $\{\mathcal{E}_q\}_{q=1}^{N_{\mathrm{E}}}$; set $\mathcal{V} \leftarrow \mathcal{V}^{(0)}$, $\mathcal{G} \leftarrow \mathcal{G}^{(0)}$, $r \leftarrow 0$, $(S^*, \hat{\theta}^*, \mathcal{I}^*, q^*) \leftarrow (\emptyset, \emptyset, \emptyset, \emptyset)$, and $J^* \leftarrow +\infty$.
\FOR{each stage $d \in \{\mathrm{sm}, \mathrm{lg}\}$ in order}
    \STATE Reset score history, stagnation base $r_{\mathrm{stag}} \leftarrow r$, and confirmation counter $c_d \leftarrow 0$.
    \IF{$d = \mathrm{lg}$}
        \STATE Set $S^* \leftarrow S_{\mathrm{sm}}^*$, retain $(\hat{\theta}^*, \mathcal{I}^*, q^*)$, and evaluate $J^{(\mathrm{lg})}(S^*)$ to initialize $J^*$.
    \ENDIF
    \FOR{$\mathrm{step} = 1, \dots, R_{\max}^{(d)}$}
        \STATE Increment $r \leftarrow r + 1$, sample active island $q_r$, and construct prompt $\Pi^{(r)}$ with island-best $S_e \in \mathcal{E}_{q_r}$, $S^*$, and diagnostic feedback.
        \STATE Sample $n_{\mathrm{b}}$ skeletons $S_j$, estimate $(\hat{\theta}_j, \hat{\Xi}_{0,j})$ on $\mathcal{D}_{\mathrm{tr}}$, and evaluate $J^{(d)}(S_j)$ via $\mathcal{I}_j$.
        \STATE Archive evaluated batch into $\mathcal{E}_{q_r}$, update $(S^*, \hat{\theta}^*, \mathcal{I}^*, q^*, J^*)$, and set $b_r \leftarrow 0$.
        \IF{$J^* \le \tau_J^{(d)}$ \textbf{and} $S^*$ satisfies stage-$d$ diagnostic criteria}
            \STATE Set $b_r \leftarrow 1$.
        \ELSIF{$S^* \neq \emptyset \wedge r - r_{\mathrm{stag}} \ge R \wedge \max_{l \in \{r-R+1, \dots, r\}} \Delta J_l \le \tau_{\Delta J}$}
            \IF{$d = \mathrm{sm} \wedge I_{\mathrm{cl}}(S^*) > \tau_{\mathrm{cl}}$}
                \STATE Update $\mathcal{V} \leftarrow \mathcal{V} \cup \mathcal{U}_v^{(r)}$ or augment $\xi_{j,h}$; set $r_{\mathrm{stag}} \leftarrow r$.
            \ELSIF{$d = \mathrm{lg} \wedge I_{\mathrm{cw}}(S^*) > \tau_{\mathrm{cw}}$}
                \STATE Expand $\mathcal{G}$ with operators $\mathcal{O}_m$ and anti-windup laws; set $r_{\mathrm{stag}} \leftarrow r$.
            \ENDIF
        \ENDIF
        \STATE Update $c_d \leftarrow b_r(c_d + 1)$.
        \IF{$c_d = R$}
            \STATE \textbf{break}
        \ENDIF
    \ENDFOR
    \IF{$d = \mathrm{sm}$}
        \STATE Lock unsaturated baseline model $S_{\mathrm{sm}}^* \leftarrow S^*$.
    \ENDIF
\ENDFOR
\RETURN $(S^*, \hat{\theta}^*, \mathcal{I}^*)$
\end{algorithmic}
\end{algorithm}

\section{Experimental Validation}

\subsection{Experimental Setup}

\begin{figure}[ht]
\centering
\includegraphics[width=1\linewidth]{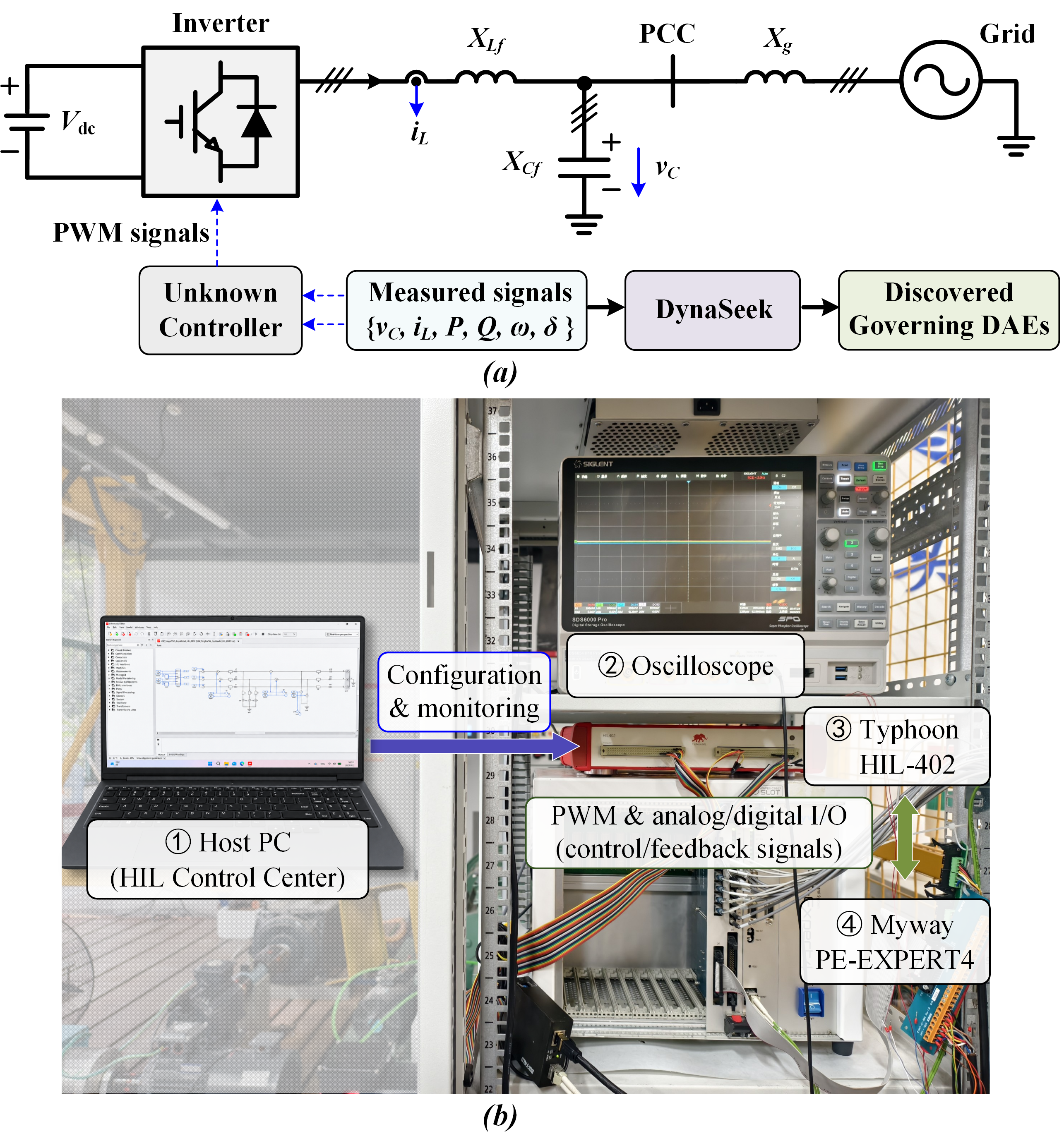}
\caption{CHIL experimental platform and measurement boundary. (a) Signal interface and accessible terminal measurements for model discovery. (b) Physical testbed comprising the host PC, Typhoon HIL-402 simulator, Myway PE-EXPERT4 controller, and oscilloscope.}
\label{fig:case_platform}
\end{figure}

To evaluate the proposed DynaSeek framework, a $100\text{-kVA}$ GFM inverter is tested on the CHIL platform shown in Fig.~\ref{fig:case_platform}, with reference system and controller parameters listed in Table~\ref{tab:case_parameters}. The power stage is emulated on a Typhoon HIL-402 simulator with a time step of $10\,\text{\textmu}\mathrm{s}$, whereas the digital controller is executed on a Myway PE-EXPERT4 platform with measurements sampled at $10\text{ kHz}$. For governing-equation discovery, DynaSeek is implemented in Python using DeepSeek-V4-Pro~\cite{xu2026deepseek} as the modeling agent, configured with the search hyperparameters and acceptance thresholds listed in Table~\ref{tab:dynaseek_configuration}.

\begin{table}[htbp]
  \centering
  \footnotesize
  \setlength{\tabcolsep}{4pt}
  \renewcommand{\arraystretch}{0.7}
  \vspace{-1em}
  \caption{Parameters of the Grid-Forming Inverter}
  \begin{tabular}{l@{\hspace{3pt}}c r l}
    \toprule
    \textbf{Parameter} & \textbf{Symbol} & \textbf{Value} & \textbf{Unit} \\
    \midrule
    \multicolumn{4}{l}{\textit{System Ratings}} \\
    Base Power & $S_{\mathrm{b}}$ & 100 & kVA \\
    Base Phase Voltage (rms) & $V_{\mathrm{n}}$ & 220 & V \\
    Nominal Frequency & $f_{\mathrm{n}}$ & 50 & Hz \\
    \midrule
    \multicolumn{4}{l}{\textit{Filter and grid parameters}} \\
    Inverter filter inductance & $L_{\mathrm{f}}$ & 0.346 & p.u. \\
    Inverter filter resistance & $R_{\mathrm{Lf}}$ & 0.035 & p.u. \\
    Inverter filter capacitance & $C_{\mathrm{f}}$ & 0.023 & p.u. \\
    Capacitor-branch damping resistance & $R_{\mathrm{Cf}}$ & 0.551 & p.u. \\
    Grid inductance & $L_{\mathrm{g}}$ & 0.199 & p.u. \\
    Grid resistance & $R_{\mathrm{Lg}}$ & 0.020 & p.u. \\
    \midrule
    \multicolumn{4}{l}{\textit{Operating references}} \\
    Active-power reference & $P^\star$ & 0.80 & p.u. \\
    Reactive-power reference & $Q^\star$ & 0 & p.u. \\
    Voltage reference & $V^\star$ & 1.008 & p.u. \\
    \midrule
    \multicolumn{4}{l}{\textit{Power-control parameters}} \\
    Inertia constant & $H$ & 0.150 & s \\
    Active-power--frequency droop coefficient & $m_{\mathrm{p}}$ & 0.075 & p.u. \\
    Reactive-power--voltage droop coefficient & $n_{\mathrm{q}}$ & 0.100 & p.u. \\
    Reactive-power--voltage loop time constant & $\tau_{\mathrm{v}}$ & 22.12 & ms \\
    \midrule
    \multicolumn{4}{l}{\textit{Voltage and current controllers}} \\
    Voltage proportional gain & $k_{\mathrm{pv}}$ & 5.5 &  \\
    Voltage integral gain & $k_{\mathrm{iv}}$ & 1450 & $\mathrm{s^{-1}}$ \\
    Current proportional gain & $k_{\mathrm{pc}}$ & 7.6 &  \\
    Current integral gain & $k_{\mathrm{ic}}$ & 11.7 & $\mathrm{s^{-1}}$ \\
    \midrule
    \multicolumn{4}{l}{\textit{Limits and anti-windup}} \\
    Current-reference magnitude limit & $I_{\max}$ & 1.8 & p.u. \\
    Internal-voltage upper bound & $E_{\max}$ & 1.15 & p.u. \\
    Voltage-controller anti-windup gain & $k_{\mathrm{aw},\mathrm{v}}$ & 1000 & $\mathrm{s^{-1}}$ \\
    Internal-voltage anti-windup gain & $k_{\mathrm{aw},E}$ & 100 & $\mathrm{s^{-1}}$ \\
    \bottomrule
  \end{tabular}
  \label{tab:case_parameters}
\end{table}

\begin{table}[ht]
    \caption{DynaSeek Search Parameters and Acceptance Thresholds}
    \label{tab:dynaseek_configuration}
    \centering
    \footnotesize
    \setlength{\tabcolsep}{4pt}
    \renewcommand{\arraystretch}{0.92}
    \begin{tabular}{p{4.8cm} >{\centering\arraybackslash}p{1.1cm} >{\centering\arraybackslash}p{1.1cm}}
        \toprule
        \textbf{Parameter} & \textbf{Symbol} & \textbf{Value} \\
        \midrule
        Experience islands & $N_{\mathrm{E}}$ & $8$ \\
        Candidates per round & $n_{\mathrm{b}}$ & $8$ \\
        Generation temperature & $T$ & $0.65$ \\
        Island-selection temperature & $\tau_{\mathrm{E}}$ & $1.0$ \\
        Stagnation detection updates & $R$ & $3$ \\
        Small-disturbance validation weights & $\alpha_{\mathrm{sm}}, \alpha_{\mathrm{lg}}$ & $1.0, 0$ \\
        Large-disturbance validation weights & $\alpha_{\mathrm{sm}}, \alpha_{\mathrm{lg}}$ & $0.2, 0.8$ \\
        Structural complexity weight & $\lambda_{\mathrm{str}}$ & $0.01$ \\
        Validation-score improvement threshold & $\tau_{\Delta J}$ & $10^{-3}$ \\
        Small-disturbance score threshold & $\tau_J^{(\mathrm{sm})}$ & $0.06$ \\
        Large-disturbance score threshold & $\tau_J^{(\mathrm{lg})}$ & $0.12$ \\
        State-closure threshold & $\tau_{\mathrm{cl}}$ & $10^{-3}$ \\
        Cross-window consistency threshold & $\tau_{\mathrm{cw}}$ & $0.1$ \\
        \bottomrule
    \end{tabular}
\end{table}

Under the prescribed information boundary, the accessible electrical measurements comprise the filter capacitor voltages $v_{\mathrm{Cd}}$ and $v_{\mathrm{Cq}}$, filter inductor currents $i_{\mathrm{Ld}}$ and $i_{\mathrm{Lq}}$, and terminal active and reactive powers $P$ and $Q$. Together with the internal angular frequency $\omega$ and power angle $\delta$, these quantities form the eight-channel measured output vector $\boldsymbol{y}$. For forward simulation, the recorded output currents $i_{\mathrm{od}}$ and $i_{\mathrm{oq}}$ and the grid angular frequency $\omega_{\mathrm{g}}$ constitute the exogenous input vector $\boldsymbol{u}$. Throughout the discovery process, internal controller states, intermediate control commands, limiter activation flags, and reference controller parameters remain strictly inaccessible.

To acquire transient records across both unsaturated and current-limited operating modes, symmetrical three-phase voltage sags lasting $150\text{--}500\text{ ms}$ are applied with depths of $5\%\text{--}20\%$ for unsaturated operation and $40\%\text{--}60\%$ for current-limited operation. Across these two operating conditions, the training set $\mathcal{D}_{\mathrm{tr}}$ comprises six unsaturated and eight current-limited transient records, while the validation sets $\mathcal{D}_{\mathrm{sm,val}}$ and $\mathcal{D}_{\mathrm{lg,val}}$ contain three and four records, respectively. For independent testing, six additional records per operating mode are reserved, covering unseen sag depths and durations excluded from the discovery process. 
From each $3\text{-s}$ recording, a transient window is extracted comprising a $100\text{-ms}$ pre-disturbance prefix, the complete voltage sag, and a $1\text{-s}$ post-fault recovery interval. For each candidate structure, model parameters are initialized via physical regression and bounded nonlinear least squares with a Soft-L1 loss, and subsequently refined by bounded coordinate search alternating with latent initial-state inference through the initialization operator $\mathcal{I}_j$ over the $100\text{-ms}$ pre-disturbance prefix. Autonomous forward trajectories are then simulated via fourth-order Runge--Kutta integration at a $20\text{-}\mu\mathrm{s}$ time step, while the normalization matrix $D_y$ remains fixed at its training-set value across validation and testing to evaluate trajectory errors on a consistent scale.

For the cross-record state-closure index $I_{\mathrm{cl}}$, each grid-voltage magnitude and phase channel is excited by $32$ nonoverlapping sine-squared probes over a $300\text{-ms}$ pre-disturbance history, with nominal probe amplitudes of $0.2\%$ for voltage magnitude and $0.02^\circ$ for phase. Following the probing interval, an identical external input trajectory is applied across all trials, and local sensitivities are estimated using multi-amplitude centered finite differences with four forward samples spaced at $0.1\text{-ms}$ intervals. For the cross-window consistency index $I_{\mathrm{cw}}$, each record in $\mathcal{D}_{\mathrm{lg,val}}$ is partitioned into overlapping evaluation windows spaced 100 ms apart, with each window spanning a 200-ms evaluation horizon sampled at 0.1-ms intervals.

\subsection{Hybrid Dynamic Discovery via the Proposed Framework}
\label{subsec:model_discovery}

\begin{figure}[ht]
\centering
\includegraphics[width=1\linewidth]{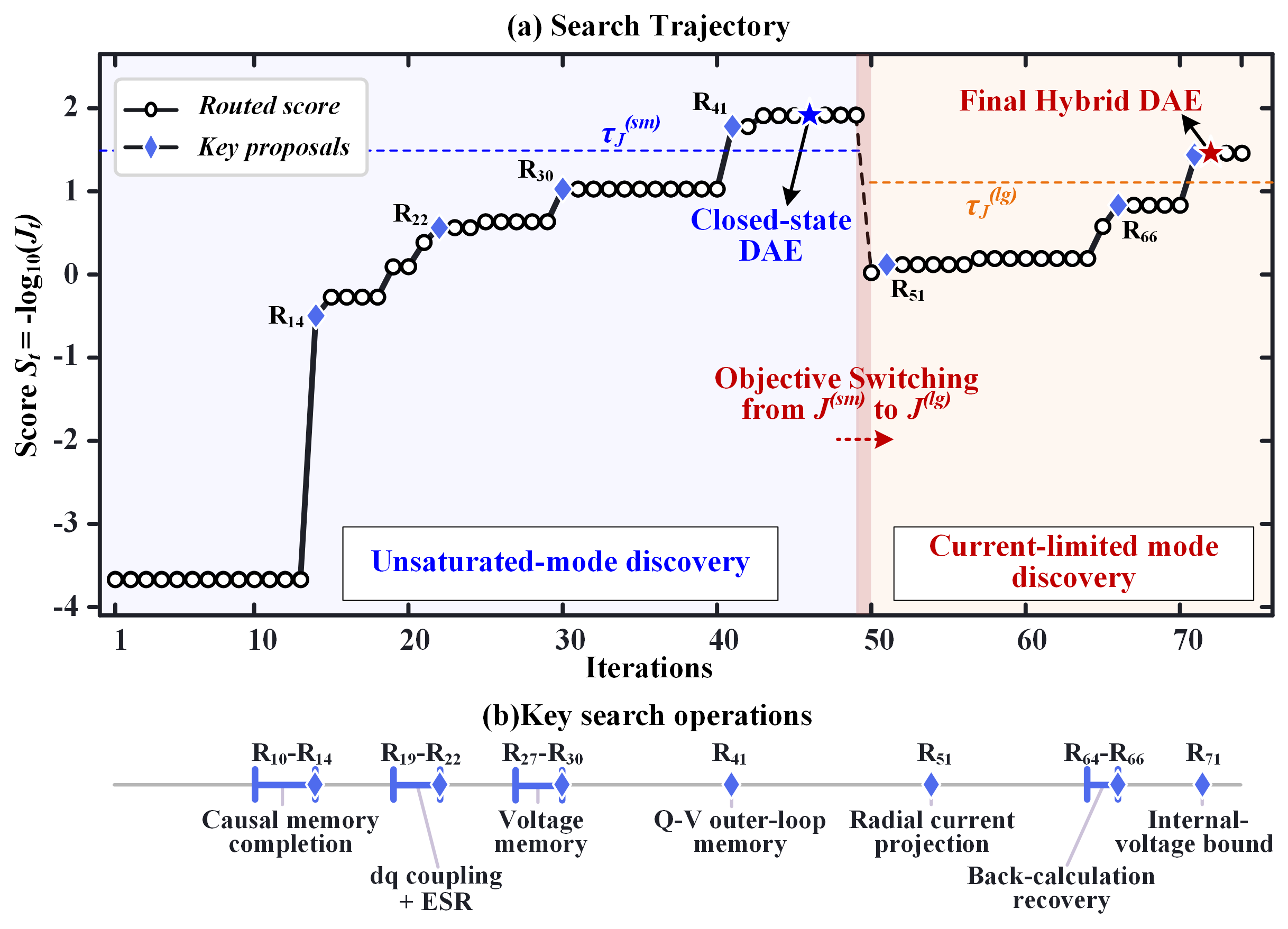}
\caption{Multi-round model discovery process. (a) Routed validation score trajectory across the unsaturated and large-disturbance stages. (b) Key structural refinements and their active search intervals.}
\label{fig:discovery_trajectory}
\end{figure}

\begin{figure}[ht]
\centering
\includegraphics[width=1\linewidth]{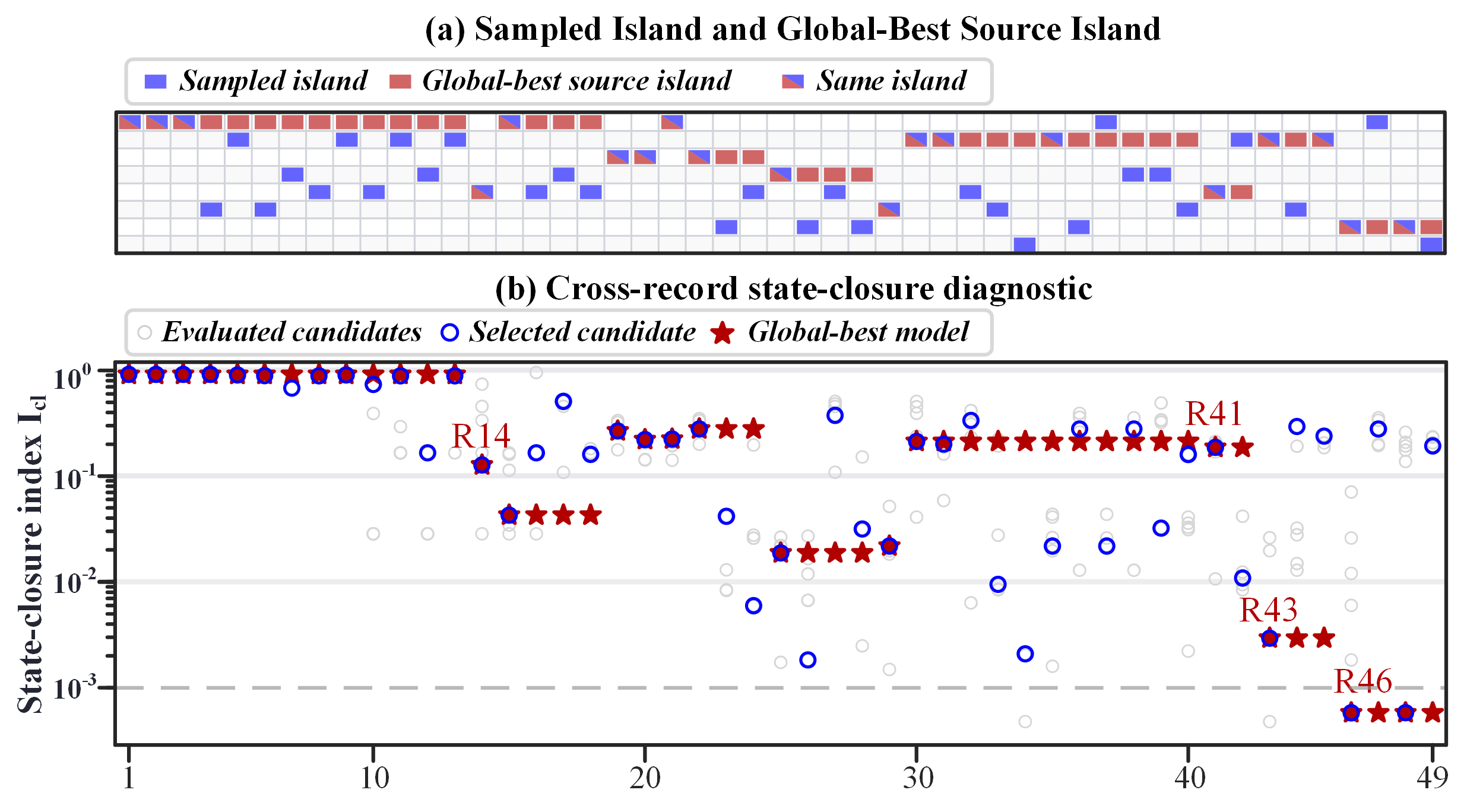}
\caption{Experience-island and state-closure evolution during unsaturated-mode discovery. (a) Sampled island and source island of the global-best model. (b) Cross-record state-closure index $I_{\mathrm{cl}}$ relative to the threshold $\tau_{\mathrm{cl}}$.}
\label{fig:Icl_discovery_trajectory}
\end{figure}

\begin{figure}[h]
\centering
\includegraphics[width=1\linewidth]{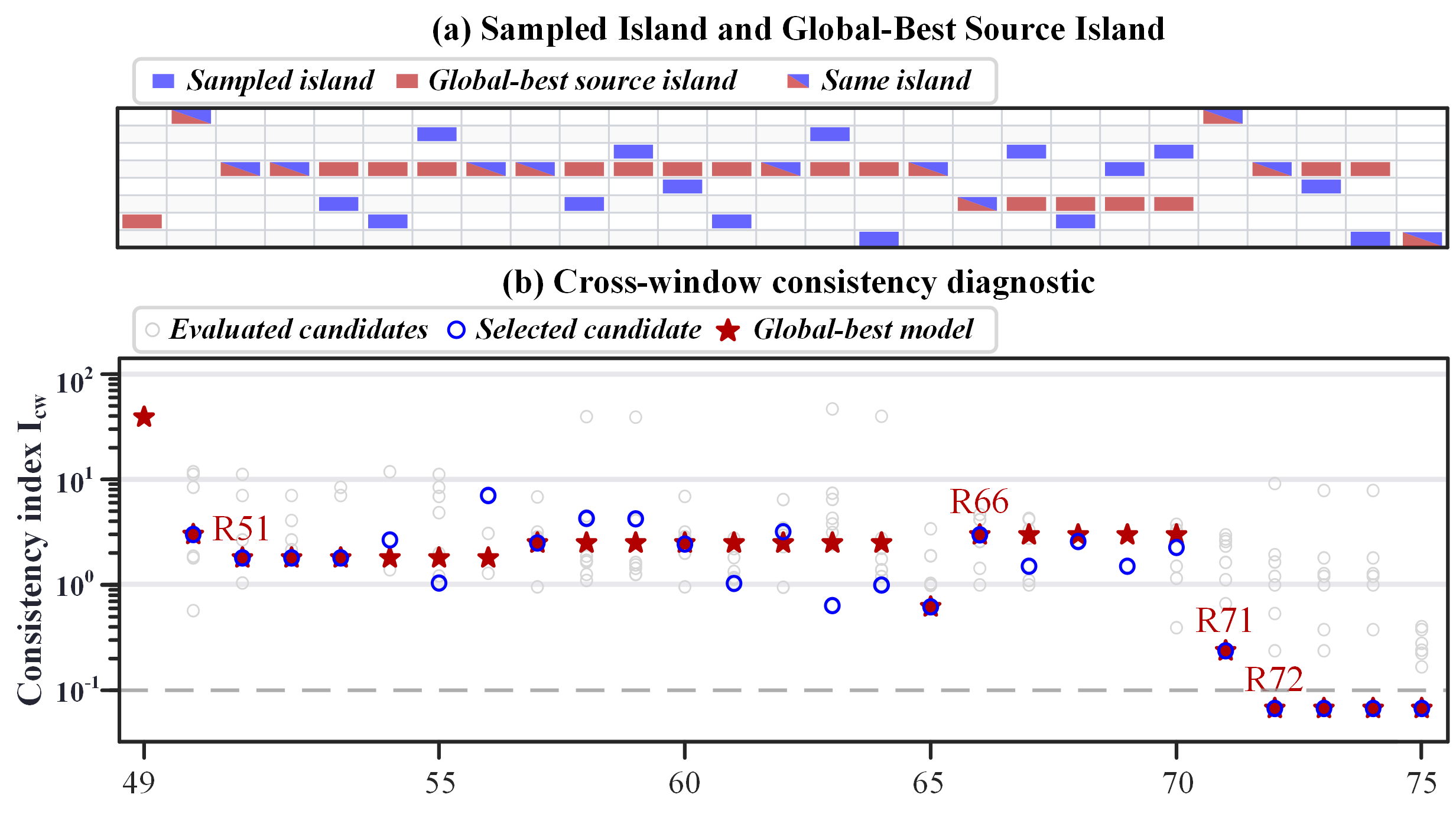}
\caption{Experience-island and consistency evolution during large-disturbance discovery. (a) Sampled island and source island of the global-best model. (b) Cross-window consistency index $I_{\mathrm{cw}}$ relative to the threshold $\tau_{\mathrm{cw}}$.}
\label{fig:Ins_discovery_trajectory}
\end{figure}

Starting from the filter circuit relations, DynaSeek first identifies the unsaturated control loops and internal control memory, as illustrated by the search trajectory in Fig.~\ref{fig:discovery_trajectory} and the state-closure evolution in Fig.~\ref{fig:Icl_discovery_trajectory}. As the composite objective $J$ penalizes both trajectory validation error and symbolic complexity, lower values of $J$ correspond to higher routed scores $S_J = -\log_{10} J$. 
Over R10--R41, the evolutionary search progressively reconstructs active-power droop synchronization, filter $dq$-axis coupling, inner-voltage regulation, and $Q$--$V$ outer-loop control, as shown in Fig.~\ref{fig:discovery_trajectory}(b). 
Although introducing the outer loop at R41 drives $J^{(\mathrm{sm})}$ past its target threshold $\tau_J^{(\mathrm{sm})}$, $I_{\mathrm{cl}}$ in Fig.~\ref{fig:Icl_discovery_trajectory}(b) remains above $\tau_{\mathrm{cl}}$ and blocks premature convergence due to the unmodeled current-loop integrator. Guided by this diagnostic violation, latent-state augmentations over R43--R46 recover the inner current-loop integral memory, reducing $I_{\mathrm{cl}}$ below $\tau_{\mathrm{cl}}$ and locking the closed-state baseline model at R49 after three confirmation rounds.

Exposing this locked baseline model to deep voltage sags at R50 switches the evaluation objective to $J^{(\mathrm{lg})}$, where unmodeled overcurrent limiting causes the sharp score drop in Fig.~\ref{fig:discovery_trajectory}(a). Through structural exploration across R51--R71, the framework incorporates a radial-projection current limiter, back-calculation anti-windup feedback, and an internal-voltage magnitude bound, bringing $J^{(\mathrm{lg})}$ within $\tau_J^{(\mathrm{lg})}$ at R71. Nevertheless, the cross-window consistency index $I_{\mathrm{cw}}$ in Fig.~\ref{fig:Ins_discovery_trajectory}(b) still exceeds $\tau_{\mathrm{cw}}$ at R71, exposing localized mode-transition mismatches between saturation entry and post-fault release along single-initial-state trajectories. Subsequent refinements of the anti-windup coupling gains over R72--R75 drive both $J^{(\mathrm{lg})}$ and $I_{\mathrm{cw}}$ within their prescribed bounds over three confirmation rounds.

\begin{table}[t]
\caption{Discovered Hybrid DAE and Its Physical Interpretation}
\label{tab:identified_equations}
\centering
\scriptsize

\setlength{\tabcolsep}{0pt}
\setlength{\arraycolsep}{0pt}
\setlength{\extrarowheight}{0pt}
\setlength{\aboverulesep}{0.8pt}
\setlength{\belowrulesep}{0.8pt}
\renewcommand{\arraystretch}{1}

\newcommand{\daeGroup}[1]{%
  \parbox[c]{32pt}{\centering\bfseries #1\par}%
}

\newcommand{\daeRole}[1]{%
  \parbox[c]{52pt}{\centering #1\par}%
}

\newcommand{\daeEq}[1]{%
  \parbox[c]{\dimexpr\columnwidth-90pt\relax}{%
    \raggedright
    \(\begin{array}{@{}>{\displaystyle}l@{}}
      #1
    \end{array}\)\par
  }\ignorespaces%
}

\newcommand{\daeRows}[1]{%
  \begin{tabular}[c]{@{}l@{\hspace{3pt}}l@{}}
    #1
  \end{tabular}%
}

\begin{threeparttable}
\begin{tabular}{@{}l@{\hspace{3pt}}l@{}}
\toprule
\daeGroup{} &
\daeRows{
  \daeRole{\textbf{Physical interpretation}} &
  \parbox[c]{\dimexpr\columnwidth-90pt\relax}{%
    \centering\bfseries Governing equations\par
  }%
}\\
\midrule

\daeGroup{Measured-state dynamics} &
\daeRows{
  \daeRole{Inductor current} &
  \daeEq{
    \dot{\boldsymbol i}_{\mathrm{L}}
    =\theta_{1,1}(\boldsymbol\xi_4-\boldsymbol v_{\mathrm{C}})
    -\theta_{1,2}\boldsymbol i_{\mathrm{L}}\\
    \quad+\omega[i_{\mathrm{Lq}},-i_{\mathrm{Ld}}]^{\mathsf{T}}.
  }\\
  \cmidrule(l){1-2}

  \daeRole{Capacitor voltage} &
  \daeEq{
    \dot{\boldsymbol v}_{\mathrm{C}}
    =\theta_{2,1}(\boldsymbol\xi_4-\boldsymbol v_{\mathrm{C}})
    +\theta_{2,2}\boldsymbol i_{\mathrm{L}}
    -\theta_{2,3}\boldsymbol i_{\mathrm{o}}\\
    \quad+\omega[v_{\mathrm{Cq}},-v_{\mathrm{Cd}}]^{\mathsf{T}}
    +\theta_{2,4}\omega[i_{\mathrm{oq}},-i_{\mathrm{od}}]^{\mathsf{T}}
    -\theta_{2,5}\dot{\boldsymbol i}_{\mathrm{o}}.
  }\\
  \cmidrule(l){1-2}

  \daeRole{Synchronization} &
  \daeEq{
    \dot{\omega}
    =-\theta_{3,1}(P-P_0)
    -\theta_{3,2}(\omega-\omega_0),\\
    \dot{\delta}=\omega-\omega_{\mathrm g}.
  }
}\\
\midrule

\daeGroup{Latent-state dynamics} &
\daeRows{
  \daeRole{Q--V control\newline \& anti-windup} &
  \daeEq{
    \dot{\xi}_1
    =-\theta_{4,1}(Q-Q_0)
    -\theta_{4,2}(v_{\mathrm{Cd}}-V_{\mathrm{d}0})\\
    \quad+\theta_{4,3}
    +\theta_{4,4}(z_1-\xi_1).
  }\\
  \cmidrule(l){1-2}

  \daeRole{Voltage loop\newline \& anti-windup} &
  \daeEq{
    \dot{\boldsymbol\xi}_2
    =\theta_{5,1}\boldsymbol z_2
    -\theta_{5,2}\boldsymbol\xi_2\\
    \quad+\operatorname{diag}(\theta_{5,3},\theta_{5,4})
    (\boldsymbol z_4-\boldsymbol z_3).
  }\\
  \cmidrule(l){1-2}

  \daeRole{Current loop} &
  \daeEq{
    \dot{\boldsymbol\xi}_3
    =\theta_{6,1}(\boldsymbol z_4-\boldsymbol i_{\mathrm{L}})
    +\theta_{6,2}[-\xi_{3,\mathrm{q}},\xi_{3,\mathrm{d}}]^{\mathsf{T}}.
  }\\
  \cmidrule(l){1-2}

  \daeRole{Arm voltage} &
  \daeEq{
    \dot{\boldsymbol\xi}_4
    =\theta_{7,1}\bigl[
    \theta_{7,2}(\boldsymbol z_4-\boldsymbol i_{\mathrm{L}})
    +\boldsymbol\xi_3-\boldsymbol\xi_4\\
    \quad+\theta_{7,3}[-i_{\mathrm{Lq}},i_{\mathrm{Ld}}]^{\mathsf{T}}\bigr].
  }
}\\
\midrule

\daeGroup{Algebraic relations} &
\daeRows{
  \daeRole{Voltage limit} &
  \daeEq{
    z_1=\min(\xi_1,\theta_{8,1}).
  }\\
  \cmidrule(l){1-2}

  \daeRole{Current\\reference} &
  \daeEq{
    \boldsymbol z_2=[z_1-v_{\mathrm{Cd}},-v_{\mathrm{Cq}}]^{\mathsf{T}},\\
    \boldsymbol z_3
    =\theta_{8,2}\boldsymbol z_2+\boldsymbol\xi_2
    +\theta_{8,3}[-v_{\mathrm{Cq}},v_{\mathrm{Cd}}]^{\mathsf{T}}.
  }\\
  \cmidrule(l){1-2}

  \daeRole{Radial limiter} &
  \daeEq{
    \boldsymbol z_4
    =\theta_{8,4}\boldsymbol z_3
    \big/\max\{\theta_{8,4},\|\boldsymbol z_3\|_2\}.
  }\\
  \cmidrule(l){1-2}

  \daeRole{Power outputs} &
  \daeEq{
    P=\theta_{8,5}(v_{\mathrm{Cd}}i_{\mathrm{Ld}}+v_{\mathrm{Cq}}i_{\mathrm{Lq}}),\\
    Q=\theta_{8,5}(v_{\mathrm{Cq}}i_{\mathrm{Ld}}-v_{\mathrm{Cd}}i_{\mathrm{Lq}}).
  }
}\\
\bottomrule
\end{tabular}

\begin{tablenotes}[flushleft]
\scriptsize
\setlength{\labelsep}{0pt}
\item[]
With reference to the control structure, $\xi_1,z_1\text{ and }\boldsymbol z_3$,$\boldsymbol z_4$ correspond to voltage and current references before and after limiting; $\boldsymbol\xi_2,\boldsymbol\xi_3$ to voltage- and current-loop memory; and $\boldsymbol\xi_4,\boldsymbol z_2$ to modulated voltage and voltage error. $\theta$ denotes model coefficients in SI units, and bold symbols denote $\mathrm{dq}$ vectors.
\end{tablenotes}

\end{threeparttable}
\end{table}

Across both stages, the frequent shifts between the sampled island and the global-best source island in Figs.~\ref{fig:Icl_discovery_trajectory}(a) and~\ref{fig:Ins_discovery_trajectory}(a) confirm that multi-island experience management continuously recombines complementary control building blocks across parallel branches. This evolutionary search ultimately yields the explicit hybrid DAE in Table~\ref{tab:identified_equations}, recovering the complete cascaded GFM control architecture without internal controller measurements. In particular, the identified internal-voltage upper bound $\theta_{8,1} = 355.165~\mathrm{V}$ and radial current limit $\theta_{8,4} = 390.286~\mathrm{A}$ closely match the $1.15\text{-p.u.}$ and $1.8\text{-p.u.}$ reference limits in Table~\ref{tab:case_parameters}, while the back-calculation terms $z_1 - \xi_1$ and $\boldsymbol{z}_4 - \boldsymbol{z}_3$ explicitly capture constraint propagation across the cascaded loops.

\begin{figure}[t]
\centering
\includegraphics[width=1\linewidth]{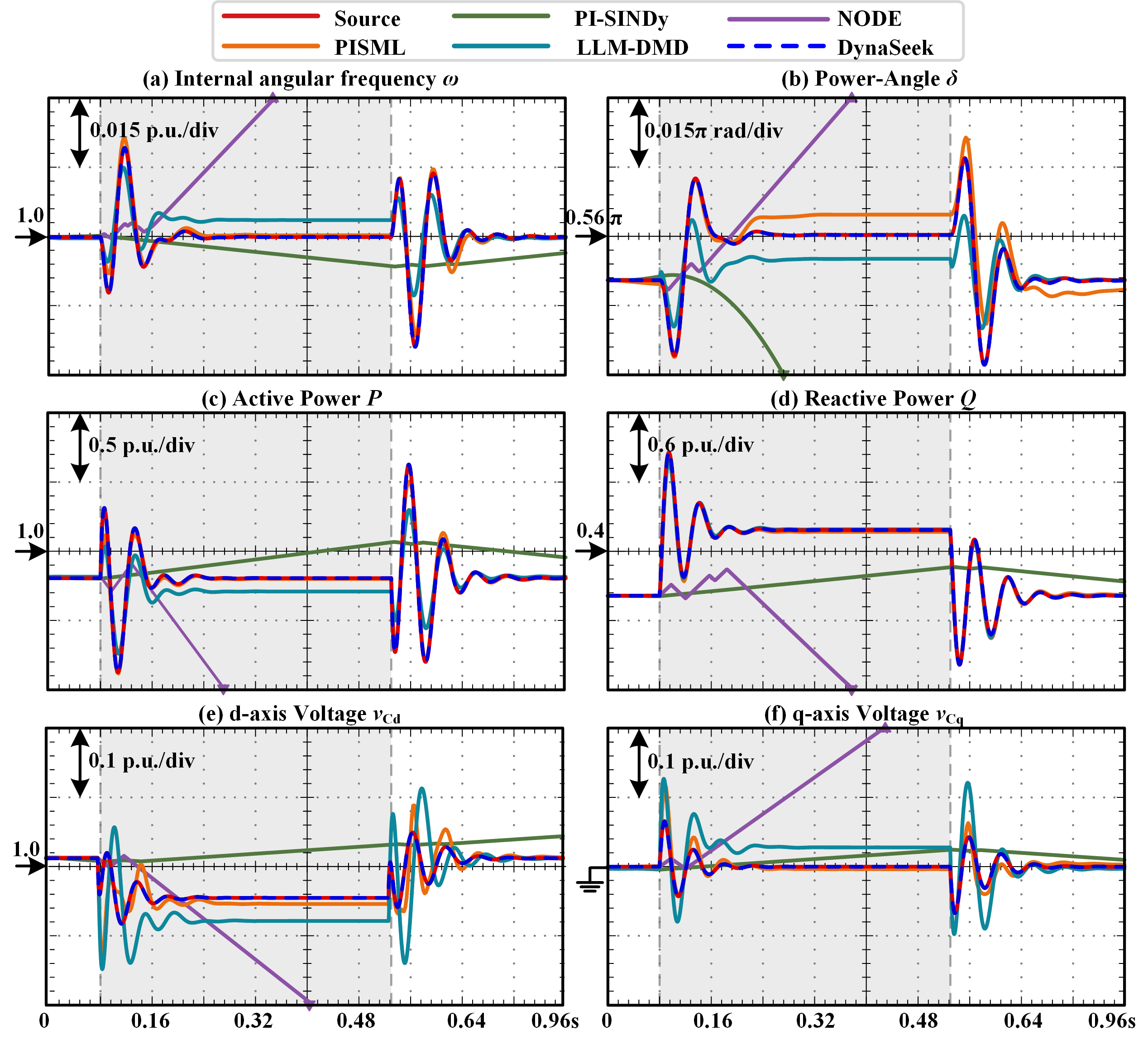}
\caption{Transient response comparison on an unseen $17.5\%$ voltage sag lasting $450\text{ ms}$ under unsaturated operation.}
\label{fig:smooth_scope}
\vspace{-1.5em}
\end{figure}

\subsection{Transient Response Reconstruction and Comparison with Baselines}

To evaluate autonomous generalization on unseen disturbances, the transient responses predicted by DynaSeek on independent test records are benchmarked against PI-SINDy~\cite{Saric2021PowerSystemSINDy}, NODE~\cite{Chen2018NeuralODE}, PISML~\cite{Zheng2026PISML}, and LLM-DMD~\cite{Shen2026LLMDMD}.
All models infer initial states solely from the 100-ms pre-disturbance window and integrate forward autonomously across both small voltage sags within the unsaturated operating range and deep voltage sags that trigger current limiting.

For the unseen 17.5\% voltage sag lasting 450 ms in Fig.~\ref{fig:smooth_scope}, the inverter operates within its unsaturated range with the current limiter inactive. Formulated solely within the measured terminal state space, NODE diverges rapidly across all channels after fault inception, while PI-SINDy exhibits monotonic drift and fails to capture the post-fault oscillatory ringdown. Although PISML and LLM-DMD improve waveform tracking through neural residuals and generated symbolic terms, lacking explicit latent control states causes persistent sag-period offsets and phase discrepancies in the power angle $\delta$ and filter capacitor voltages $v_{\mathrm{Cd}}$ and $v_{\mathrm{Cq}}$. By recovering the underlying latent control states, DynaSeek accurately captures the cumulative dynamics of internal regulation errors, closely tracking the reference transient peaks, damped oscillations, and steady-state operating point across all channels.

\begin{figure}[t]
\centering
\includegraphics[width=1\linewidth]{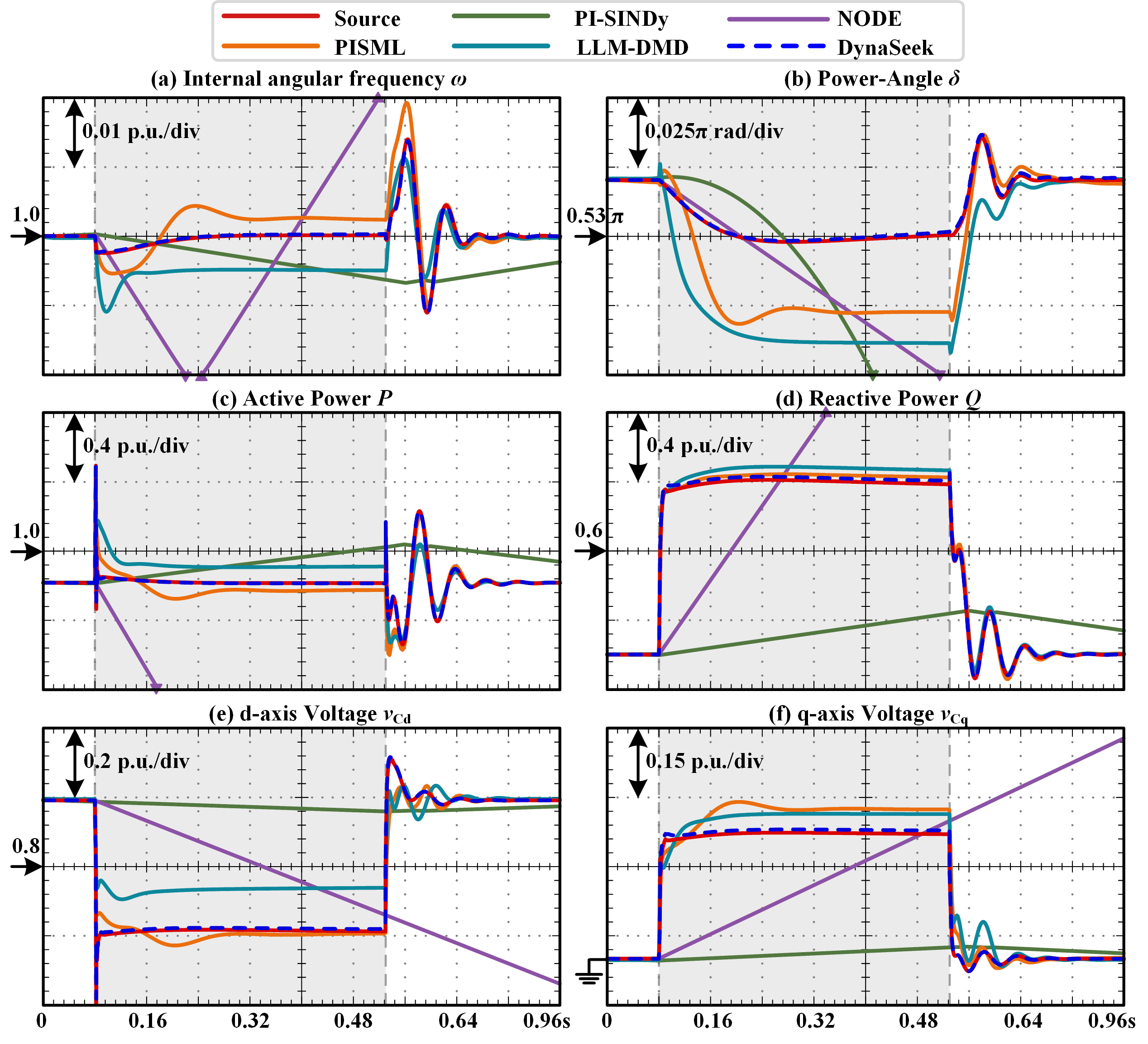}
\caption{Transient response comparison on an unseen $57.5\%$ voltage sag lasting $450\text{ ms}$ with current-limit activation.}
\label{fig:limit_scope}
\vspace{-1em}
\end{figure}

When the disturbance deepens to an unseen 57.5\% voltage sag lasting 450 ms in Fig.~\ref{fig:limit_scope}, the inverter enters current-limited operation immediately upon fault inception. Clamping the current reference reconfigures the active regulation loops and alters the internal integrator states that govern post-fault recovery. While the measured-state-only baselines again diverge and fail to reproduce either the clamped plateau or the post-fault ringdown, the residual-assisted and LLM-based baselines attempt to absorb these saturation-induced mode transitions into continuous nonlinear terms. During the clamped interval, this structural deficiency induces spurious frequency excursions in $\omega$, pronounced steady-state deviations in $P$ and $v_{Cd}$, and severe negative power-angle drift in $\delta$, followed by distorted recovery overshoots and phase lags after fault clearance. In contrast, DynaSeek couples the radial-projection current limiter and back-calculation anti-windup laws with the latent control states, accurately reproducing both the clamped fault response and the post-fault recovery trajectory across all measured channels.

\begin{figure}[t]
\centering
\includegraphics[width=1\linewidth]{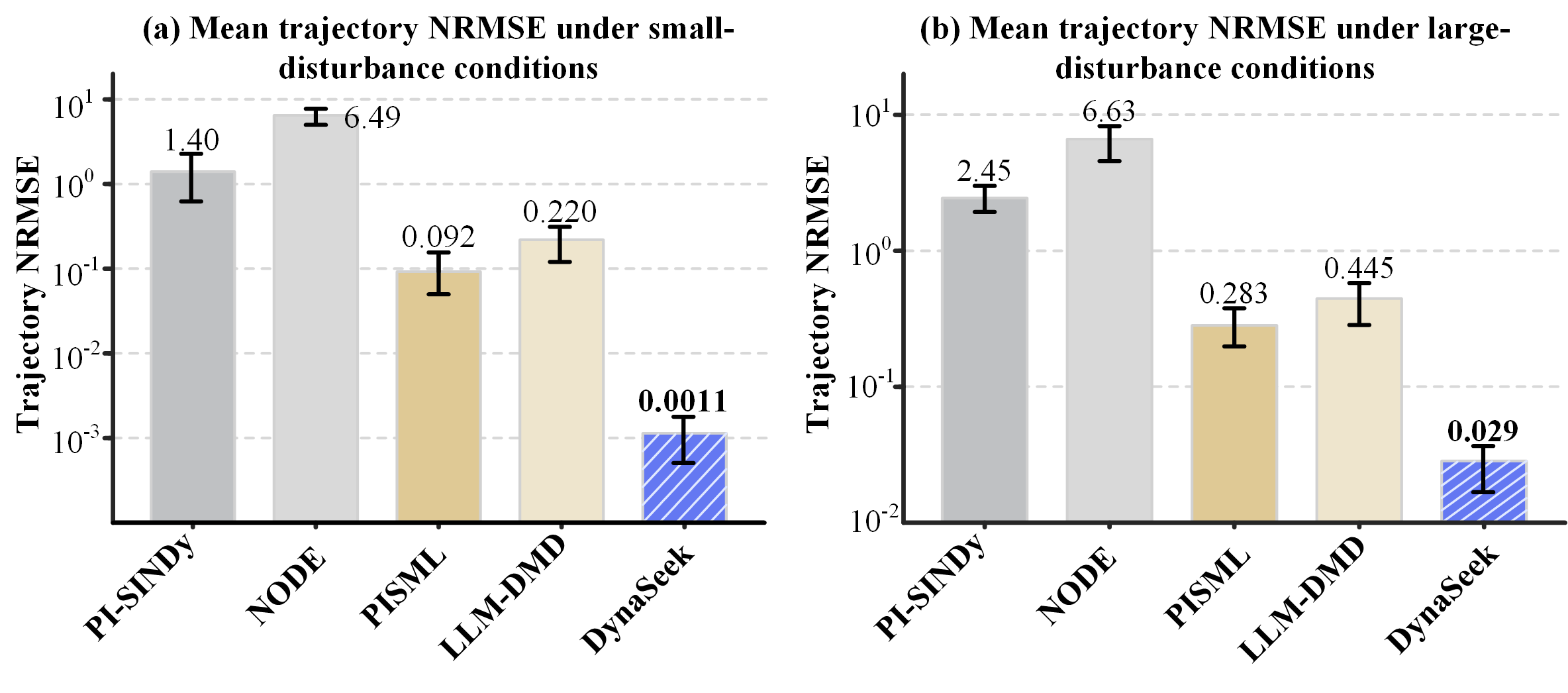}
\caption{Trajectory NRMSE comparison across six random seeds under (a) small-disturbance and (b) large-disturbance test conditions, with error bars showing the minimum-to-maximum range.}
\label{fig:seeds}
\vspace{-1.5em}
\end{figure}

Fig.~\ref{fig:seeds} summarizes the trajectory normalized root-mean-square errors (NRMSEs) evaluated over the independent test set across six random seeds. DynaSeek achieves mean NRMSEs of 0.0011 under small disturbances and 0.029 under large disturbances, reducing prediction errors by nearly two orders and one order of magnitude relative to the strongest baseline PISML, respectively, and by even wider margins relative to the remaining baselines. Although trajectory sensitivity naturally increases around current-limiting boundaries, the narrow minimum-to-maximum error bars across all six independent runs confirm that the diagnosis-guided multi-island search consistently converges to high-fidelity hybrid DAEs despite the stochastic nature of LLM sampling.

\subsection{System-Level Integration and Stability-Oriented Assessment}

\begin{figure}[ht]
\centering
\includegraphics[width=0.85\linewidth]{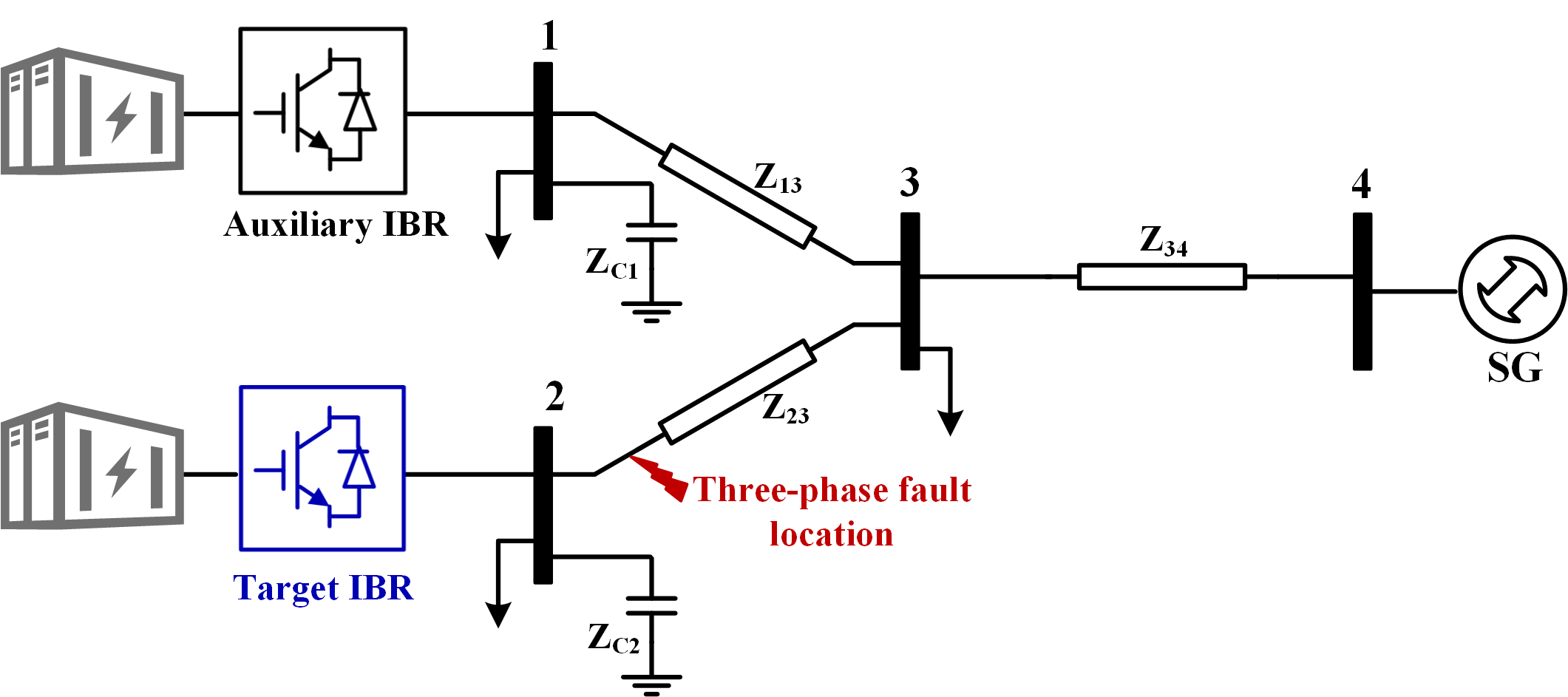}
\caption{Modified four-bus test system for closed-loop transient stability and CCT evaluation.}
\label{fig:4BusSystem}
\end{figure}

For network-level transient stability assessment, the identified hybrid DAE is embedded into the modified four-bus test system shown in Fig.~\ref{fig:4BusSystem}, which comprises two GFM inverters and one synchronous generator. Under identical network parameters and pre-fault operating conditions, the target IBR at Bus 2 is represented by either the reference white-box model or the discovered hybrid DAE, whereas the auxiliary IBR at Bus 1 and the synchronous generator at Bus 4 remain unchanged. Interfaced through terminal voltage and current variables, the discovered model interacts with the external network equations in a fully autonomous closed-loop manner.

Within the four-bus system, grid strength is varied from weak to strong grids by adjusting line impedances, while moderate to severe three-phase faults at Bus 2 are generated by varying the fault resistance. Across all test conditions in Table~\ref{tab:cct_comparison}, DynaSeek correctly identifies every stable operating case and reproduces all finite ground-truth CCTs within 0.1 ms. In contrast, both PISML and LLM-DMD suffer from qualitative stability misclassifications and substantial quantitative errors. Under moderate sags where the reference system remains stable, these baselines introduce false-instability predictions. For deeper faults with finite stability boundaries, they systematically underestimate the CCT, collapsing to a small fraction of the ground truth in extreme cases. Consistent with the large-disturbance responses in Fig.~\ref{fig:limit_scope}, trajectory deviations in active power and power angle during current limiting distort the clearing state, whereas an incomplete characterization of internal integrator windup and recovery shrinks the post-fault attraction domain.

The ablation results in Table~\ref{tab:cct_comparison} further isolate the distinct roles of the two structural diagnostics. Without state-closure guidance, DynaSeek w/o $I_\mathrm{cl}$ converges to an incomplete latent-state structure lacking essential integral memory, yielding a false-instability error under moderate sag conditions alongside persistent CCT underestimation under deeper faults. Disabling cross-window consistency diagnosis produces an even more hazardous failure mode. Unable to recover explicit current-limiting and anti-windup laws, DynaSeek w/o $I_\mathrm{cw}$ misses the finite stability boundary under the strong-grid moderate sag, erroneously classifying an unstable fault as stable while severely underestimating the CCT under deep sags. These contrasting failures confirm that neglecting either internal control memory or saturation-induced mode transitions compromises reliable stability-boundary assessment.

\begin{table}[t]
\caption{Critical Clearing Times (ms) Under Different Grid Strengths and Voltage-Sag Depths}
\label{tab:cct_comparison}
\centering
\scriptsize
\setlength{\tabcolsep}{1.6pt}
\renewcommand{\arraystretch}{1.12}

\begin{tabular}{@{}cccccccc@{}}
\toprule
\multicolumn{2}{c}{\textbf{Test Conditions}}
& \multicolumn{6}{c}{\textbf{CCT (ms)}} \\

\cmidrule(lr){1-2}
\cmidrule(lr){3-8}

\makecell[c]{\textbf{SCR}}
& \makecell[c]{\textbf{Sag depth}\\\textbf{(\%)}}
& \makecell[c]{\textbf{Ground}\\\textbf{truth}}
& \makecell[c]{\textbf{PISML}}
& \makecell[c]{\textbf{LLM-DMD}}
& \makecell[c]{\textbf{DynaSeek}\\\textbf{w/o }$I_{\mathrm{cl}}$}
& \makecell[c]{\textbf{DynaSeek}\\\textbf{w/o }$I_{\mathrm{cw}}$}
& \makecell[c]{\textbf{DynaSeek}} \\

\midrule

\multirow[c]{3}{*}{2}
& 40 & Stable & Stable & 114.7 & Stable & Stable & \textbf{Stable} \\
& 60 & 125.7 & 102.5 & 66.9 & 102.6 & 57.2 & \textbf{125.7} \\
& 80 & 69.8 & 63.2 & 19.1 & 63.3 & 16.4 & \textbf{69.7} \\

\midrule

\multirow[c]{3}{*}{3}
& 40 & Stable & 143.4 & 56.2 & 200.1 & Stable & \textbf{Stable} \\
& 60 & 80.6 & 64.0 & 19.8 & 67.5 & 29.6 & \textbf{80.6} \\
& 80 & 51.4 & 44.7 & 5.4 & 44.8 & 13.1 & \textbf{51.4} \\

\midrule

\multirow[c]{3}{*}{5}
& 40 & 157.6 & 74.3 & 20.1 & 101.0 & Stable & \textbf{157.5} \\
& 60 & 54.9 & 45.0 & 9.2 & 48.3 & 17.5 & \textbf{54.9} \\
& 80 & 38.9 & 35.7 & 2.9 & 35.7 & 10.2 & \textbf{38.9} \\

\bottomrule
\end{tabular}

\vspace{3pt}
\begin{minipage}{\columnwidth}
\footnotesize
\raggedright
Note: w/o $I_{\mathrm{cl}}$ and w/o $I_{\mathrm{cw}}$ denote models discovered without
state-closure and cross-window consistency diagnostics, respectively.
Stable indicates stability for all tested fault durations
up to 2 s.
\end{minipage}
\vspace{-2em}
\end{table}

\begin{figure}[ht]
\centering
\includegraphics[width=1\linewidth]{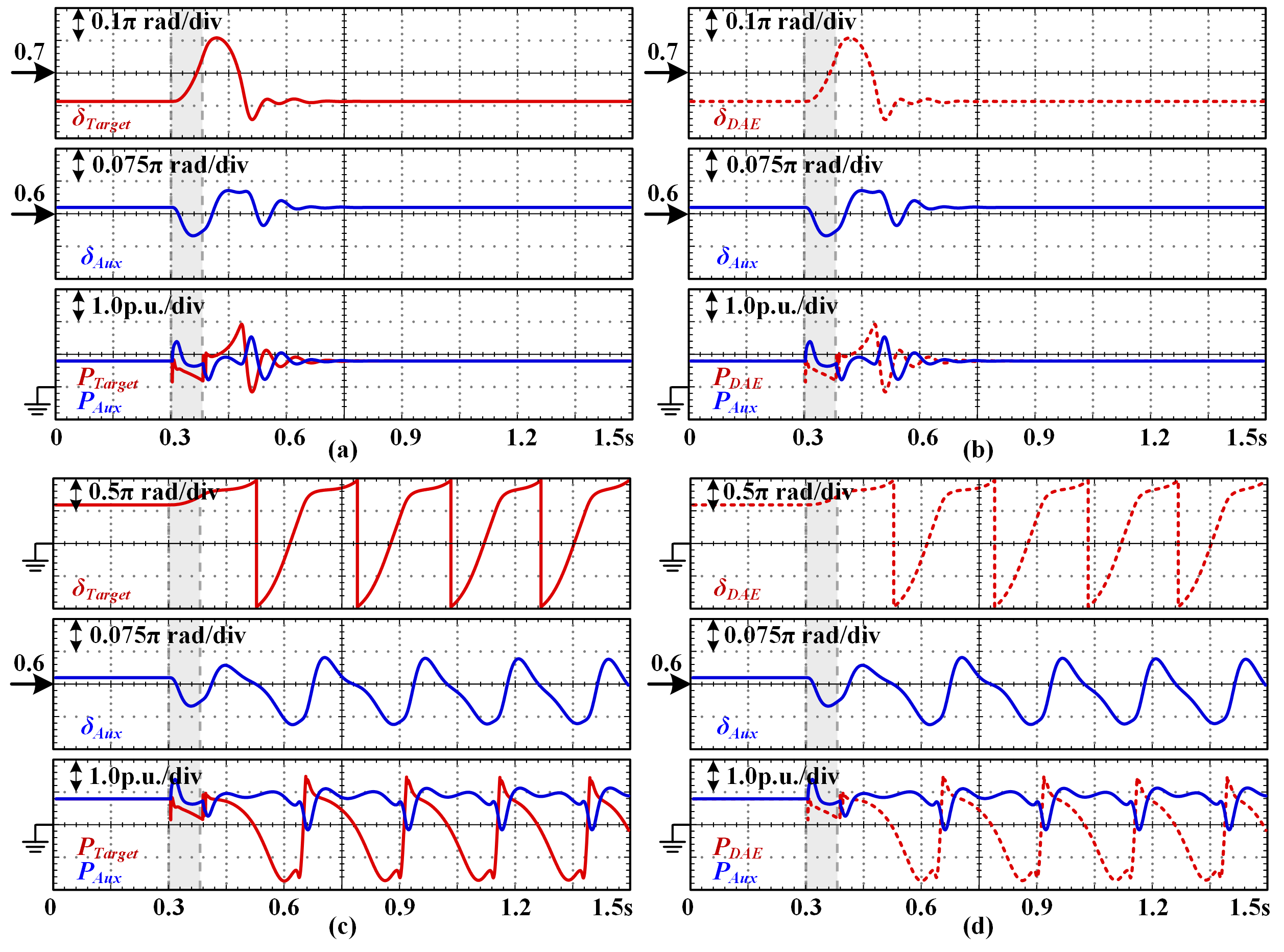}
\caption{Closed-loop transient responses near the ground-truth CCT under $\mathrm{SCR} = 3$ and a $60\%$ voltage sag. (a), (c) Reference system cleared $1\text{ ms}$ below and $1\text{ ms}$ above the ground-truth CCT. (b), (d) Corresponding responses with the Target IBR represented by the discovered hybrid DAE.}
\label{fig:CCT_compare}
\end{figure}

To inspect closed-loop dynamic behavior near the stability boundary, Fig. 12 compares the post-fault responses under SCR = 3 and a 60\% voltage sag for clearing times set 1 ms below and 1 ms above the ground-truth CCT. When the fault is cleared 1 ms below the CCT, the power angles remain bounded and the active-power oscillations damp out in Figs.~\ref{fig:CCT_compare}(a) and (b), confirming that both the reference system and the DynaSeek-embedded system maintain transient synchronism. Conversely, when the clearing time exceeds the CCT by 1 ms, both systems exhibit loss of synchronism at the target IBR with periodic angle wrapping and sustained post-fault power oscillations in Figs.~\ref{fig:CCT_compare}(c) and (d). 
Furthermore, the transient trajectories of the auxiliary IBR at Bus 1 closely align between the two systems, confirming accurate capture of inter-inverter network coupling. Overall, these closed-loop evaluations verify that the discovered hybrid DAE preserves the large-signal and stability-oriented dynamics of the proprietary inverter for self-contained system-level simulation.

\section{Conclusion}

This paper has presented DynaSeek, an LLM-driven framework for discovering explicit hybrid DAE models of grid-tied inverters from limited measurements. Formulating governing-equation discovery as a diagnosis-guided progressive structural search, the framework integrates joint parameter and initial-state estimation with cross-record state-closure and cross-window consistency diagnostics to direct latent control-state augmentation and current-limiting and anti-windup refinement. Multi-island experience management preserves diverse candidate structures across parallel evolutionary branches, while staged validation maintains unsaturated modeling accuracy as the baseline model expands to current-limited operation. CHIL experiments on a 100-kVA GFM inverter demonstrate that DynaSeek reconstructs both unsaturated and current-limited transient responses with trajectory errors one to two orders of magnitude lower than existing symbolic and neural baselines. When integrated into a modified four-bus power system for autonomous closed-loop simulation, the discovered hybrid DAE accurately reproduces post-fault trajectories and reliably predicts reference CCTs across diverse grid strengths and fault severities, while ablation studies confirm the necessity of retaining both latent control memory and current-limiting and anti-windup dynamics. Future work will extend this diagnosis-guided discovery paradigm to inverters with diverse control-mode transitions and aggregated dynamic modeling of multi-inverter renewable power plants.

\bibliographystyle{IEEEtran}
\bibliography{mybibfile}

\end{spacing}
\end{document}